\documentclass[twocolumn,showpacs,preprintnumbers,amsmath,amssymb,floatfix,prd]{revtex4}

\usepackage{epsfig}

\begin{document}

\title{Determination of pion and kaon fragmentation
 functions including spin asymmetries data in a global analysis}

\author{M. Soleymaninia$^{a,b}$}
\email{Maryam_soleymaninia@ipm.ir}

\author{A. N. Khorramian$^{a,b,c}$}
\email{Khorramiana@theory.ipm.ac.ir}

\author{S. M. Moosavi Nejad$^{d,b}$}
\email{Mmoosavi@yazduni.ac.ir}

\author{F. Arbabifar$^{a,b}$}
\email{Farbabifar@ipm.ir}

\affiliation{$^{(a)}$ Faculty of Physics, Semnan University, 35131-19111 Semnan, Iran \\
$^{(b)}$ School of Particles and Accelerators, Institute for Research
in Fundamental Sciences (IPM), P.O.Box 19395-5531, Tehran, Iran\\
$^{(c)}$Stanford Institute for Theoretical Physics and Department of Physics, Stanford University, Stanford, CA 94305-4060, USA\\
$^{(d)}$Faculty of Physics, Yazd University, P.O.Box 89195-741, Yazd, Iran\\
}

\date{\today}
\begin{abstract}
We present new functional form of pion and kaon fragmentation functions up to next-to-leading order
obtained through a global fit to single-inclusive electron-positron annihilation data and also
employ, the semi-inclusive deep inelastic scattering asymmetry data from
HERMES and COMPASS to determine fragmentation functions. Also we apply very recently electron-positron annihilation data from BaBar and Belle at
$\sqrt{s}=10.54$~GeV and $\sqrt{s}=10.52$~GeV, respectively.
In this analysis we consider the impression of semi-inclusive deep inelastic scattering asymmetry
data on the fragmentation functions, where the produced hadrons of different electric charge are identified. We break symmetry assumption between quark and anti-quark fragmentation functions for favored partons by using the asymmetry data. The results of our analysis are in good agreement with electron-positron annihilation data and also with all the semi-inclusive deep inelastic scattering asymmetry data.
Also we apply the obtained fragmentation functions
to predict the scaled-energy distribution of $\pi^+/K^+$ inclusively produced in top-quark decays
at next-to-leading order using the zero-mass variable-flavor-number scheme exploiting the universality and
scaling violations of fragmentation functions.
\end{abstract}


\maketitle
\section{Introduction}
In high energy processes at Relativistic Heavy Ion Collider (RHIC) and Large Hadron Collider (LHC), the quantum chromodynamics
(QCD) predictions of cross sections have very important role. In the general case the
parton distribution functions (PDFs) of initial hadrons,  parton-level differential cross sections,
and fragmentation functions are three necessary ingredients
to calculate cross sections. Fragmentation functions (FFs) form  the non-perturbative component of
the hard scattering process and describe the transition of a parton into  the outgoing hadron.\\
The importance of FFs is for model independent predictions of the cross
sections at LHC in which a hadron is detected in the outgoing
productions depend on FFs. Interest of FFs is found for example in tests of QCD such
as theoretical calculations for recent measurements of inclusive production in proton-proton
collisions at RHIC, and in investigating the  origin of
the proton spin. In the naive parton model the FFs are independent of the factorization scale
(depending only on the scaling variable $z$) but in QCD-improved parton model the
scaling violations of FFs remain to be subject to Dokshitzer-
Gribov-Lipatov-Alteralli-Parisi (DGLAP) evolution equations \cite{dglap}.
Note that FFs can be extracted from fits to data at intermediate to large momentum fractions.\\
FFs are studied in electron-positron annihilation, lepton-hadron and hadron-hadron scattering processes.
Among all, the $e^+e^-$ annihilation provides a clean environment to determine the fragmentation
densities, but without an initial hadron state one can not separate a quark from an anti-quark distributions.
Since  the most precise data from $e^+e^-$ annihilation
exists for the production of the  lightest charged hadrons (pion, kaon and proton), we are interested  in the
fragmentation processes of the partons into the pion and kaon in electron-positron
annihilation and semi-inclusive deep inelastic scattering. Fragmentation functions are included in hadron-production
processes in electron-positron annihilation, lepton-proton or nucleus scattering, proton-proton and heavy
ions collisions. Such processes are important in hadron physics for studying and investigating
the origin of the proton spin \cite{deFlorian:2005mw,deFlorian:2009vb} and in reported results in Refs.~\cite{Kretzer:2002qh,Kretzer:2001pz,Artru:2010st,Artru:1998sb}, the role of FFs to determine
the polarized parton distribution functions is pointed out. In the present analysis to determine FFs,
we consider semi-inclusive deep inelastic scattering (SIDIS) asymmetry data along with the data on the $e^+e^-$ annihilation
from  LEP (ALEPH, DELPHI and OPAL collaborations), SLAC (SLD and TPC
collaborations), DESY (TASSO collaboration), KEK (TOPAZ collaboration)
\cite{aleph91,delphi91,delphi91-2,opal91,sld91,tasso34_44,tpc29,topaz58} and very recent BaBar and
Belle data at the SLAC and the KEKB \cite{Lees:2013rqd,Leitgab:2013qh}, respectively.
There are also some other kinds of new analysis performed by using these new reported data \cite{Anderle:2013pla,Jain:2012uq}.\\
There are already several theoretical studies on QCD analysis for FFs which are listed in \cite{hira}, in particular,
used parameterizations were widely obtained by KKP and AKK collaborations \cite{Kniehl} and
Kretzer \cite{Kretzer:2000yf}. As  it is shown in Ref.~\cite{Hirai:2007cx}, there are
differences between the fragmentation functions of KKP and Kretzer, therefore it is convenient to make an error calculation
for the fragmentation functions to check the consistency of results. In this regard, an attempt for determining the
FFs and their uncertainties has been already done in HKNS and DSS collaborations~\cite{deFlorian:2007aj,Hirai:2007cx}.
But we have an opportunity to use the asymmetry SIDIS experimental data from HERMES \cite{Hermes05} and COMPASS \cite{Alekseev:2010ub,Alekseev:2009ac} which is the first attempt
to study how much the asymmetry SIDIS data are effective for determination of FFs and also their uncertainties.\\
Since the decay of top-quark is one of the interesting subjects at LHC we shall make our theoretical predictions for
the energy spectrum of  $\pi^+$- and $K^+$-mesons produced in top decay using the FFs obtained from
our present work. We will also compare our predictions with the known results of Refs.~\cite{Albino:2008fy,Hirai:2007cx,deFlorian:2007aj}.\\
This paper is organized as follows. In Section~\ref{sec:two} we explain the hadronization process in electron-positron
collision by introducing the FFs. Double spin asymmetry in semi-inclusive deep inelastic scattering
processes is defined in this section too. In section~\ref{sec:three} we describe our formalism and
parametrization form for pion and kaon fragmentation densities. The global $\chi^2$ minimization for data is defined in this section.
We also outline the Hessian method for assessing the neighborhood of global minimum. In section~\ref{sec:four} our formalism to predict
the energy distribution of  $\pi^+$ and $K^+$ in top-quark decay is explained.
The full results for pion and kaon FFs and their uncertainties are listed in section~\ref{sec:five}. Comparison of our results with experimental data and the other models and also our predictions of energy spectrum of outgoing pion and kaon in top-quark decay are presented in this section.
Our conclusion is given in section~\ref{sec:six}.

\section{Theoretical formalism for fragmentation functions}
\label{sec:two}
The fragmentation functions are related to the low energy components of the hadron production processes
and they form the non-perturbative aspect of QCD.  FFs describe the inclusive emission of a hadron from
a parton and they cannot be precisely calculated by theoretical methods at this stage. The perturbative QCD framework
is used to study single-inclusive hadron production in $e^+e^-$ annihilation, lepton-nucleon DIS, and hadron-hadron
collisions, where the factorization theorem is a strong tool to study such processes. This theorem states that the cross section
can be expressed in terms of  perturbatively calculable partonic hard-scattering
cross sections, PDFs and FFs in which the two last are related to the
low  energy components of QCD processes. The low energy components of QCD processes are universal and they can be used to make predictions.\\
In this section, we explain the theoretical framework relevant for our global QCD analysis
of fragmentation functions.
\subsection{Single-inclusive $e^+e^-$ annihilation}
Since, in the single-inclusive $e^{+}e^{-}$ annihilation processes
\begin{eqnarray}\label{single}
e^{+}e^{-}\rightarrow (\gamma ,Z)\rightarrow H+X,
\end{eqnarray}
one should not deal with the uncertainty introduced by PDFs in comparison with the hadron collisions, then
the optimal way to determine FFs is to fit them to experimental date extracted from these processes.
In the above process,  $X$ stands for  the unobserved jets which are produced along with a
detected hadron $H$.  \\
The perturbative QCD framework is used to study hadron production
in $e^{+}e^{-}$ annihilation, where the factorization theorem is
an important tool to study this process.
According to the hard-scattering factorization theorem
of the parton model \cite{esw-book,Collins:1998rz}, the cross section can be written
as a sum of convolutions of perturbatively calculable the partonic hard scattering cross
sections $d\sigma_a(y, \mu_R, \mu_F)/dy$ \cite{Kretzer:2000yf,Kniehl:2005de,Binnewies:1994ju} with the
non-perturbative fragmentation functions of the hadron $H$ from a parton $a$, $D^H_a (x, \mu_F)$ as following
\begin{eqnarray}
\label{sigx}
\frac{1}{\sigma_{tot}}\,\frac{d\sigma^{H}}{dz}=
\sum_a\int\limits_z^1\frac{dy}{y}\,D_a^{H}(\frac{z}{y}, \mu_F)\,
\frac{1}{\sigma_{tot}}\,
\frac{d\sigma_a}{dy}\left(y, \mu_R,  \mu_F\right),\nonumber\\
\end{eqnarray}
where the $a$ stands for one of the partons $a=g, u,\bar u, \cdots, b,\bar b$.
We denote the four-momenta of the intermediate
gauge boson and the $H$ hadron by $q$ and $p_H$, respectively, so that $s=q^2$ and the scaling variable $z$
is defined as $z=2(p_H.q)/q^2$. In the center-of-mass (c.m.) frame, $z$ is simplified to  $z=2E_H/\sqrt{s}$ which refers
to the energy of $H$ scaled to the beam energy.
The function $D_a^H(x,\mu_F)$ indicates the probability to find the hadron $H$ from a parton $a$  with
the scaled energy fraction $x$. In equation above, $y$ is defined in analogy to $z$ as $y=2(p_a . q)/q^2$, where $p_a$ is the
four-momentum of parton  $a$.
The renormalization and factorization scales are given by $\mu_R$ and $\mu_F$, respectively, however
one can choose two different values for them but a choice often made consists of setting $\mu_R^2=\mu_F^2=Q^2$
and we shall adopt this convention in this work. \\
At NLO, the total cross section  is described by the $q\bar q$-pair creation subprocesses
, $e^+e^-\rightarrow (\gamma, Z)\rightarrow q\bar q+(g)$,  as
\begin{eqnarray}
\sigma_{tot}=N_c \sigma_0\sum_{i=1}^{n_f}(V^2_{q_i}+A^2_{q_i})\big[1+
\frac{\alpha_s(\mu)}{2\pi}C_F\frac{3}{2}+{\cal O}(\alpha^2_s)\big],\nonumber\\
\end{eqnarray}
where $C_F=(N_c^2-1)/(2N_c)=4/3$ for $N_c=3$ quark colors and $\sigma_0=(4\pi\alpha^2/3s)$ is the leading order total cross
section of $e^+e^-\to\mu^+\mu^-$ for massless muons,
 $n_f$ is the number of active flavors, $N_c=3$ is the number of quark colors and
 $\alpha$ is the electro-weak coupling constant.
 $V_{q_i}$ and $A_{q_i}$ are the effective vector and axial-vector couplings of quark $q_i$ to the both
 intermediate  photon and $Z$-boson, which  can be found in Ref.~\cite{Kneesch:2007ey}.
 For small energies, $\sqrt{s}\ll M_Z$, for the  summation of squared effective electro-weak charges one has
 $V^2_{q_i}+A^2_{q_i}=e^2_{q_i}$ where $e_{q_i}$ is the electric charge of quark $q_i$, see Ref.~\cite{Kneesch:2007ey}.\\
In our global analysis of FFs, some of the data sets from the OPAL and the ALEPH
experiments are in the form of $1/N_{Z\rightarrow Hadron} (d N^H/d z)$
 where $N$ is the number of detected events. This  is defined as the ratio of   the single-inclusive
 $e^{+}e^{-}$ annihilation  cross section (\ref{sigx})  in a certain bin of $z$ to the totally inclusive  rate, i.e.
\begin{equation}
\frac{1}{N_{tot}} \frac{d N^H}{d z} \equiv
\frac{1}{\sigma_{tot}} \frac{d \sigma^H}{dz}.
\end{equation}

\subsection{Hadronization process in $ep$ collisions and spin asymmetry}
\label{sec:three}
Generally parton distribution function $q(x,Q^2)$ expresses the probability density to find a parton $q$ in a nucleon carrying fraction $x$ of target nucleon
momentum at the transfer momentum $Q^2$.
 For the polarized PDFs, we assume that a proton is made
of massless partons with the positive and negative helicity
distributions and thus the difference
\begin{equation}
\delta q(x, Q^2) = q_{+}(x, Q^2) - q_{-}(x, Q^2),
\end{equation}
demonstrates how much the parton of flavor $q$
represents of the proton polarization.
Universally, the functional forms of the polarized and unpolarized
PDFs are determined by a QCD fit to the experimental data obtained
from various interactions. These analysis have been discussed in
lots of recent reviews
\cite{Gluck:1998xa,Pisano:2008ms,Gluck:2006pm,Goto:1999by,deFlorian:2005mw,deFlorian:2008mr,Hirai:2008aj,Blumlein:2010rn,Leader:2010dx,Khorramian:2004ih,Atashbar
Tehrani:2007be,deFlorian:2009vb,lss,PRD11,AtashbarTehrani:2011zz,Arbabifar:2011zz,Khanpour:2010zz,TaheriMonfared:2011zz,Khorramian:2010qa,Khanpour:2012tk,TaheriMonfared:2010zz}
and new more precise investigations are still
in progress. \\
In order to cover more kinematics region in the current analysis we
consider polarized SIDIS process, $\vec{l}(l) + \vec{N}(p_N)
\rightarrow l^{'}(l^{'}) + H(p_H) +X$, where hadron $H$ is also
detected along with the scattered lepton $l^{'}$ and jets  $X$. This
process gives a remarkable information concerning the nucleon
structure in quite distinct kinematics which probes different
aspects of fragmentation distributions. Moreover, SIDIS data help us
to specify the difference between the quark and anti-quark
distributions in the nucleon considering outgoing produced hadrons
 which is not
possible in fully inclusive experiments.\\
 In the polarized SIDIS, the measured double spin asymmetry $A_{1}^{N,H}$ can be expressed in terms of  the
ratio of polarized and unpolarized structure functions $g_{1}^{N,H}$
and $F_{1}^{N,H}$ \cite{lss}, as
\begin{equation}
A_{1}^{N,H}(x,z,Q^2)=\frac{g_{1}^{N,H}(x,z,Q^2)_{NLO}}{F_{1}^{N,H}(x,z,Q^2)_{NLO}},
\label{A1h}
\end{equation}
where $z$ is the scaled energy fraction of the outgoing hadron,
$Q^2$ is the transfer momentum and $x$ is the Bjorken scaling
variable. The
indices $N$ and $H$ stand for the different nucleon targets  and outgoing  detected hadron, respectively.\\
In the  NLO approximation, the polarized and unpolarized structure
functions $g_1^{N,H}$ and $F_1^{N,H}$ in SIDIS processes are presented as (see
Ref.~\cite{lss})
\begin{eqnarray}
\label{g1h}
 2g_1^{N,H}(x,z,Q^2)&=&\sum _{q,\bar{q}}
^{n_f}e_{q}^2\left\{\hspace{-0.4cm}\phantom{\int\limits_a^b}\Delta
q(x,Q^2) D_q^H(z,Q^2)\right.
\nonumber\\
&&+
\frac{\alpha_s(Q^2)}{2\pi}\left[\hspace{-0.4cm}\phantom{\int}\Delta
q\otimes \Delta C_{qq}^{(1)} \otimes D_q^H\right.
\nonumber\\
&&+ \Delta q\otimes \Delta C_{gq}^{(1)}\otimes D_g^H
\nonumber\\
&&+ \left.\left.\hspace{-0.4cm}\phantom{\int}\Delta g\otimes \Delta
C_{qg}^{(1)}\otimes
D_q^H\right](x,z,Q^2)\phantom{\int\limits_a^b}\hspace{-0.4cm}\right\},\nonumber\\
\end{eqnarray}
and
\begin{eqnarray}
2F_1^{N,H}(x,z,Q^2)&=&\sum _{q,\bar{q}}
^{n_f}e_{q}^2\left\{\hspace{-0.4cm}\phantom{\int\limits_a^b}q(x,Q^2)
D_q^H(z,Q^2)\right.
\nonumber\\
&&+
\frac{\alpha_s(Q^2)}{2\pi}\left[\hspace{-0.4cm}\phantom{\int}q\otimes
C_{qq}^{(1)}\otimes D_q^H\right.
\nonumber\\
&&+ q\otimes C_{gq}^{(1)}\otimes D_g^H
\nonumber\\
&&+ \left.\left.\hspace{-0.4cm}\phantom{\int}g\otimes
C_{qg}^{(1)}\otimes
D_q^H\right](x,z,Q^2)\phantom{\int\limits_a^b}\hspace{-0.4cm}\right\},\nonumber\\
\label{F1h}
\end{eqnarray}
where $n_f$ is the number of active flavors, $e_q$ is the electric
charge of quark $q$, $\alpha_s$ is the strong coupling constant,
$\Delta q$ and $q$
 are polarized and unpolarized parton densities and $\Delta C_{ij}^{(1)}(x,z)$ and
$C_{ij}^{(1)}(x,z)$ are the polarized and unpolarized NLO Wilson
coefficient functions, respectively, presented in Ref.~\cite{FSV}. The
corresponding parton FFs, $D_{q,\bar{q}}^H$ and $D_g^H$, are
determined in the present global analysis and play a significant
role in determination of $A_{1}^{N,H}$.\\
According to Eq.~(\ref{A1h}) the double spin asymmetry $A_{1}^{N,H}$ depends on polarized and unpolarized parton distribution functions so
in order to calculate the double spin asymmetry, we need to
use the results of available PPDFs and PDFs sets.
Here we choose the latest DSSV PPDFs and KKT12 PDFs \cite{deFlorian:2009vb,Khanpour:2012tk},
however, the different choices of PPDFs and PDFs do not change our result considerably.\\
\section{QCD Analysis \& Parametrization}
\label{sec:three}
\subsection{ZM-VFN scheme}\label{subsec:Afour}
In a parton fragmentation function $D^{H}_{i}(z,\mu^{2})$, $z$ represents the fraction
of a parton's momentum carried by a produced hadron while in a parton distribution  $q(x,\mu^2)$, $x$
represents the fraction of a hadron's momentum carried by constituent parton. In both
cases, QCD parton model approach would predict $z$ and $x$-distributions independent
of the factorization scale. Note that the similar violations of this scaling behavior are happened when QCD
corrections are taken into account \cite{esw-book}, in  other words,
beyond the leading order  of perturbative QCD these universal
functions are factorization-scale dependent.
The $z$ dependence of the fragmentation functions is of non-perturbative aspect of QCD
and  they are not yet calculable from first principles. However, once they
are given at the initial fragmentation scale $\mu_0$ their $\mu$ evolution is determined by the DGLAP
 renormalization group equations \cite{dglap} which is very similar to
 those for parton densities. For example, the flavor-singlet evolution equation reads \cite{Almasy:2011eq}
\begin{eqnarray}\label{apff}
\frac{\partial}{\partial \ln \mu^2}
\left(\begin{array}{ccc}
D_S^H(z, \mu^2)\\
D_g^H(z, \mu^2)\\
\end{array}\right)&=&
\left(\begin{array}{cccc}
P_{qq}(z) \quad   P_{gq}(z)\\
P_{qg}(z)  \quad  P_{gg}(z)\\
\end{array}\right)\nonumber\\
&&\otimes
\left(\begin{array}{ccc}
D_S^H(z, \mu^2)\\
D_g^H(z, \mu^2)\\
\end{array}\right),
\end{eqnarray}
where $D_S^H(z, \mu^2)$ refers to the singlet function, $D_S^H(z, \mu^2)=\sum_q [D_q^H(z, \mu^2)+D_{\bar q}^H(z, \mu^2)]$,
and the convolution integral $\otimes$  is defined by
\begin{eqnarray}
f(z) \otimes g(z)=\int_z^1\frac{dy}{y} f(y) g(\frac{z}{y}).
\end{eqnarray}
The functions $P_{ji}$ are time-like splitting functions which are the same as those in
deep inelastic scattering at the lowest order but  the higher order terms are different.
The third-order contributions (N$^{2}$LO) to the quark-gluon and gluon-quark timelike splitting functions
 can be found in Ref.~\cite{Almasy:2011eq}. The evolution equations are essentially
the same as the PDF case, so that the same
numerical method  can be used to obtain a solution.
Also the flavor non-singlet evolution equation can be found in Refs.~\cite{Mitov:2006ic,Albino:2010jc,Albino:2011gq}.\\
To extract the fragmentation functions from data analysis there are several approaches and in the present analysis we adapt the zero-mass variable-flavor-number (ZM-VFN) scheme \cite{jm}. In this
scheme,  all quarks are treated as massless particles and the non-zero values of the charm  and
bottom-quark masses only enter through the initial conditions of the FFs. The
 number of active flavors  also varies with the factorization scale where
 for the scales higher than the respective flavor thresholds, the quark is active as a
parton. This scheme works best for high energy scales, where $m_Q = 0$ is a good approximation.\\
We also evaluate $\alpha_s^{(n_f)}(\mu)$ at NLO in the improved-minimal subtraction ($\overline{\text{MS}}$)
scheme using
\begin{eqnarray}\label{alpha}
\alpha^{(n_f)}_s(\mu)=\frac{1}{b_0\log(\mu^2/\Lambda^2)}
\Big\{1-\frac{b_1 \log\big[\log(\mu^2/\Lambda^2)\big]}{b_0^2\log(\mu^2/\Lambda^2)}\Big\},
\nonumber\\
\end{eqnarray}
with
\begin{eqnarray}
b_0=\frac{33-2n_f}{12\pi}, \quad  b_1=\frac{153-19n_f}{24\pi^2},
\end{eqnarray}
where  $\Lambda$ is the typical QCD scale. We adopt
$\Lambda_{QCD}^{(4)}=334.0$~MeV \cite{Martin:2002aw} for NLO adjusted such
that $\alpha_s^{(5)}=0.1184$ for $M_Z=91.1876$~GeV \cite{Nakamura:2010zzi}.
Also $\Lambda_{QCD}^{(4)}$ at LO is fixed in our fits to $220.0$~MeV.

\subsection{Parametrization of fragmentation functions}
\label{subsec:threeB}
We parameterize the pion and kaon fragmentation functions at LO and NLO considering
the single inclusive annihilation (SIA) and SIDIS data.  At the initial scale $\mu_{0}$
this parametrization contains a functional form as
\begin{equation}
\label{ff-input}
D_i^H(z,\mu_{0}^{2}) =
N_i z^{\alpha_i}(1-z)^{\beta_i} [1-e^{-\gamma_i z}],
\end{equation}
which  is a convenient form for the light hadrons. A simple polynomial parametrization with
just 3 parameters $N_i$, $\alpha_i$ and $\beta_i$ controls the small and large $z$ region \cite{Soleymaninia:2012zz}.
Accordingly, power term in $z$ emphasizes the small $z$ region and power term in $(1-z)$
restricts the large $z$ region. We consider the extra term $[1-e^{-\gamma_i z}]$ to control
medium $z$ region and to improve the accuracy of the global fit \cite{Soleymaninia:2012ss}.
The free parameters $N_i$, $\alpha_i$, $\beta_i$ and $\gamma_i$ are determined by fitting $\chi^{2}$
of SIA and SIDIS data.  In the $\overline{MS}$ scheme,
there is an important sum rule of the FFs on the energy conservation as
\begin{equation}
\label{esum}
\int_0^1 dz\ z\ \sum_H\ D_i^H(z,\mu^2) = 1,
\end{equation}
which means that each parton surly will fragment into some hadron $H$. Since the summation
over all the hadrons can not be taken practically and the  behavior of small $z$ is unstable, Eq.~(\ref{esum})
can not be a viable constraint in a global analysis.\\
The initial scale $\mu_{0}$ is different for partons.
The starting  scale for the FFs of the  light-quarks
$(u/\bar{u}, d/\bar{d}, s/\bar{s})$  and $g$ into $\pi^\pm/K^\pm$-mesons is $\mu_{0}^{2}=1$~GeV$^2$
and it is taken at $\mu_{0}^{2}=m_{c}^{2}$ and $\mu_{0}^{2}=m_{b}^{2}$ for charm and bottom-quarks
\cite{Hirai:2010cs,de Florian:2007nk}. We choose $m_{c}=1.43$~GeV
and $m_{b}=4.3$~GeV in our analysis. Then these FFs are evolved to
higher scales using the DGLAP group equations in Eq.~(\ref{apff}).\\
In all analysis it is necessary to use different assumptions for various FFs.
In our analysis, we take the same fragmentation densities for valence quarks.
Since the possibility of $\pi^+/K^+$-production from valence or favored quarks is more than sea
or unfavored quarks then we assume distinct fragmentation functions for the light sea quarks.
Because of mass difference, the different functions are also specified for heavy quarks and
we assume the same FFs for heavy quark and its anti-quark like $D^{H}_{c}=D^{H}_{\bar{c}}$ and $D^{H}_{b}=D^{H}_{\bar{b}}$.
According to the pion structure $|\pi^+\rangle =| u\bar{d}\rangle$  and the general
functional form presented in  Eq.~(\ref{ff-input}), for the parton-FFs into $\pi^+$ one has
\begin{eqnarray}\label{pifit1}
& D_{u,\bar{d}}^{\pi^{+}} & (z,\mu_{0}^{2}) =N_{u}^{\pi^{+}} z^{\alpha_{u}^
{\pi^{+}}}(1-z)^{\beta_{u}^{\pi^{+}}} [1-e^{-\gamma_{u}^{\pi^{+}} z}],
\nonumber\\
& D_{d,\bar{u},s,\bar{s}}^{\pi^{+}} & (z,\mu_{0}^{2}) =N_{d}^{\pi^{+}}
z^{\alpha_{d}^{\pi^{+}}}(1-z)^{\beta_{d}^{\pi^{+}}} [1-e^{-\gamma_{d}^{\pi^{+}} z}],
\nonumber\\
\end{eqnarray}
where we impose $SU(2)$ isospin invariance between $u$ and $\bar{d}$-quarks for pion.
Isospin symmetry is also considered for sea quarks of pion fragmentation functions. Gluon and heavy quarks FFs
are defined as
\begin{eqnarray}\label{pifit2}
& D_{g}^{\pi^{+}} & (z,\mu_{0}^{2}) =N_{g}^{\pi^{+}} z^{\alpha_{g}^{\pi^{+}}}
(1-z)^{\beta_{g}^{\pi^{+}}} [1-e^{-\gamma_{g}^{\pi^{+}} z}],
\nonumber\\
& D_{c,\bar{c}}^{\pi^{+}} & (z,m_{c}^{2}) =N_{c}^{\pi^{+}} z^{\alpha_{c}^{\pi^{+}}}
(1-z)^{\beta_{c}^{\pi^{+}}}[1-e^{-\gamma_{c}^{\pi^{+}} z}],
\nonumber\\
& D_{b,\bar{b}}^{\pi^{+}} & (z,m_{b}^{2}) =N_{b}^{\pi^{+}} z^{\alpha_{b}^{\pi^{+}}}
(1-z)^{\beta_{b}^{\pi^{+}}}[1-e^{-\gamma_{b}^{\pi^{+}} z}].
\nonumber\\
\end{eqnarray}
Considering the constituent quark composition of kaon $|K^+\rangle =| u\bar{s}\rangle$,
we define kaon functional form for light partons as following
\begin{eqnarray}\label{kafit1}
& D_{u}^{K^{+}} & (z,\mu_{0}^{2}) =N_{u}^{K^{+}} z^{\alpha_{u}^{K^{+}}}
(1-z)^{\beta_{u}^{K^{+}}} [1-e^{-\gamma_{u}^{K^{+}} z}],
\nonumber\\
& D_{\bar{s}}^{K^{+}} & (z,\mu_{0}^{2}) =N_{{\bar{s}}}^{K^{+}}
z^{\alpha_{\bar{s}}^{K^{+}}}(1-z)^{\beta_{\bar{s}}^{K^{+}}}
[1-e^{-\gamma_{\bar{s}}^{K^{+}} z}],
\nonumber\\
& D_{d,\bar{d},\bar{u},s}^{K^{+}} & (z,\mu_{0}^{2}) =N_{d}^{K^{+}}
z^{\alpha_{d}^{K^{+}}}(1-z)^{\beta_{d}^{K^{+}}} [1-e^{-\gamma_{d}^{K^{+}} z}].
\nonumber\\
\end{eqnarray}
We apply a new form of kaon FF for the strange-quark because of its mass against the $u$-quark mass.
Kaon fragmentation functions for sea quarks defined by considering isospin symmetry between them.
Gluon and heavy quarks FFs are defined as following
\begin{eqnarray}\label{kafit2}
& D_{g}^{K^{+}} & (z,\mu_{0}^{2}) =N_{g}^{K^{+}} z^{\alpha_{g}^{K^{+}}}
(1-z)^{\beta_{g}^{K^{+}}} [1-e^{-\gamma_{g}^{K^{+}} z}],
\nonumber\\
& D_{c,\bar{c}}^{K^{+}} & (z,m_{c}^{2}) =N_{c}^{K^{+}} z^{\alpha_{c}^{K^{+}}}
(1-z)^{\beta_{c}^{K^{+}}},
\nonumber\\
& D_{b,\bar{b}}^{K^{+}} & (z,m_{b}^{2}) =N_{b}^{K^{+}} z^{\alpha_{b}^{K^{+}}}
(1-z)^{\beta_{b}^{K^{+}}}.
\end{eqnarray}
In our analysis, the heavy partons parameters $\gamma_c$ and $\gamma_b$ impress the $\chi^2$ value. We use Eq.~\ref{ff-input}
as a functional form for all partons of pion and our decision to put or omit the term $[1-e^{-\gamma_iz}]$
for different flavors of kaon FFs is based on getting the best $\chi^2$. In reported parameters in Refs.~\cite{Kretzer:2000yf,Hirai:2007cx,deFlorian:2007aj}
some parameters are fixed or in the other word the simple parametrization form is used.
In Ref.~\cite{Hirai:2007cx} one of the gluon parameters is fixed for pion and kaon
and DSS model \cite{deFlorian:2007aj} uses the simple parametrization for $c$ and $b$-quarks for $\pi^+$ and $K^+$ mesons.\\
According to the parton structure of $\pi^- (\bar{u}d)$ and $K^- (\bar{u}s)$, parton
fragmentation functions could be calculated for $\pi^-$ and $K^-$ as
\begin{eqnarray}
\label{minus1}
D_{i}^{\pi^{-}}(z,\mu_{0}^{2}) = D_{\bar{i}}^{\pi^{+}}(z,\mu_{0}^{2}),\\ \nonumber
D_{i}^{K^{-}}(z,\mu_{0}^{2}) = D_{\bar{i}}^{K^{+}} (z,\mu_{0}^{2}),
\end{eqnarray}
where $i=u,d,s,c,b$ and also for gluon fragmentation functions
\begin{eqnarray}
\label{minus2}
D_{g}^{\pi^{-}}(z,\mu_{0}^{2}) = D_{g}^{\pi^{+}}(z,\mu_{0}^{2}),\\ \nonumber
D_{g}^{K^{-}}(z,\mu_{0}^{2}) = D_{g}^{K^{+}}(z,\mu_{0}^{2}).
\end{eqnarray}
%
\begin{table*}[th]
\caption{\label{tab:lopionpara} Fit parameters for
the parton FFs into the  charged pion ($\pi^+$) at LO, $D_i^{\pi^+}(z,\mu_0)$.
The starting scale is taken to be $\mu_{0}^2=1$~GeV$^2$ for light partons and gluon and also $\mu_{0}^2=m_c^{2}$ and $\mu_{0}^{2}=m_b^{2}$ for $c$ and $b$-quarks.}
\begin{ruledtabular}
\begin{tabular}{cccccc}
flavor $i$ &$N_i$ & $\alpha_i$ & $\beta_i$ &$\gamma_i$\\
\hline
$u, \overline{d}$ &$ 0.841\pm 0.435$&$-2.041\pm0.417$& $0.803\pm0.284$&$1.170\pm0.789$
\\
$\overline{u}, d, s, \overline{s}$ & $2.509\pm1.143$&$-1.514\pm0.269$&$4.610\pm1.083$&$2.123\pm1.189$
\\
$c, \overline{c}$ &$1.059\pm0.788$&$-1.918\pm0.364$& $3.325\pm1.060$&$1.697\pm1.447$
\\
$b, \overline{b}$ & $1.062\pm0.596$&$-2.043\pm0.218$&$ 5.902\pm1.328$& $1.750\pm1.044$
\\
$g$              & $59.993\pm10.523$&$ 0.939\pm1.961$& $5.801\pm5.394$&$1.219\pm3.508$
\\
\end{tabular}
\end{ruledtabular}
\end{table*}
%
\begin{table*}[t]
\caption{\label{tab:nlopionpara} Values of fit parameters for the  $\pi^+$ meson at NLO
  in the starting scale.}
\begin{ruledtabular}
\begin{tabular}{cccccc}
flavor $i$ &$N_i$ & $\alpha_i$ & $\beta_i$ &$\gamma_i$\\
\hline
$u, \overline{d}$ &$ 1.049\pm0.563$&$-1.916\pm0.421$& $0.977\pm0.304$&$0.964\pm0.65$
\\
$\overline{u}, d, s, \overline{s}$ & $9.968\pm7.441$&$-0.516\pm0.481$&$5.952\pm1.565$&$1.898\pm1.885$
\\
$c, \overline{c}$ &$ 0.946\pm0.859$&$-1.723\pm0.451$& $3.590\pm1.280$&$1.947\pm1.981$
\\
$b, \overline{b}$ & $0.869\pm0.550$&$-2.059\pm0.234$&$ 5.803\pm1.460$& $1.561\pm1.054$
\\
$g$              & $219.507\pm44.789$&$ 1.073\pm0.362$& $7.505\pm1.140$&$2.142\pm1.411$
\\
\end{tabular}
\end{ruledtabular}
\end{table*}
\subsection{QCD analysis of Experimental data and global minimization of $\chi^{2}$}
\label{subsec:chi2}
The free parameters in the functional forms of  $\pi^+$ and $K^+$ FFs
( Eqs.~(\ref{pifit1}-\ref{kafit2})) are determined by minimizing $\chi^2$ for
 differential cross section and asymmetry experimental data (i.e. $(1/\sigma_{tot}\cdot d\sigma/dz)_{exp}$ and $(A_{1}^{N,H})_{exp}$)
in $x$ space. The global $\chi^2$ is defined as
\begin{equation}
\label{eq:chi2}
\chi^2_{global}=\sum_{n} w_{n}\chi ^{2}_{n},
\end{equation}
where $n$ is the number of experimental data group and $w_n$ denotes a weight factor for the $n$-th experimental data group. The $\chi ^{2}_{n}$ is defined as
\begin{equation}
\label{eq:chi2}
\chi^2_{n}=(\frac{1-N_n}{\Delta N_n})^2+\sum_{j=1}^k (\frac{N_n E_j-T_j}{N_n\sigma^{E}_{j}})^2,
\end{equation}
where $T_{j}$ and $E_j$ are the theoretical and  experimental values of
 $1/\sigma_{tot}\cdot d\sigma/dz$ for $e^+e^-$ SIA data
 and $A_{1}^{N,H}$ for SIDIS data and $\sigma^{E}_{j}$
is the error of corresponding experimental value.
Here, the summation goes over the $k$ bins of the experimental data. $\Delta N_n$ related to the experimental normalization uncertainty which is reported by the experiments and $N_n$ corresponds to an overall normalization factor which is refer to the data of experiments. Usually $N_n$ is gotten from first minimization and we fix it in the second minimization.
In our global fit we take SIA experimental data
from LEP (ALEPH, DELPHI and OPAL collaborations), SLAC (BaBar, SLD and TPC
collaborations), DESY (TASSO collaboration) and KEK (Belle and TOPAZ collaborations).
The energy scales of experimental data
are from $10.52$~GeV to $91.2$~GeV \cite{aleph91,delphi91,delphi91-2,opal91,Lees:2013rqd,sld91,tasso34_44,tpc29,Leitgab:2013qh,topaz58}.
In the reported data without discrimination of  hadron species, authors distinguished
between four cases; fragmentation of $u, d, s$-quarks, $c$-quark
only, $b$-quark, and all five quark flavors  ($u, d, s, c$, and $b$).
These categories are just in DELPHI and SLD data \cite{delphi91,delphi91-2,sld91}.\\
Also the BaBar and Belle collaborations \cite{Lees:2013rqd,Leitgab:2013qh} reported inclusive
hadron production cross sections recently at a center-of-mass energy of $10.54$~GeV and $10.52$~GeV, respectively.
Since the center-of-mass energies are below the threshold to produce
a $b\bar{b}$ pair, these data containing a pure $e^+e^- \rightarrow q\bar{q}$ sample, where
$q={u, d, s, c}$. Although most of the precision $e^+e^-$ annihilation data, are limited
to results from experiments at LEP and SLAC at the energy scale of the $M_Z$,
the large data samples are available at BaBar and Belle collaborations at $Q=10.54$~GeV and $Q=10.52$~GeV,
respectively. In addition, these two collaborations reported the differential cross sections at $z > 0.7$.\\
We also provide
SIDIS experimental data for $A_{1}^{p,\pi ^+}$, $A_{1}^{p,\pi ^-}$,
$A_{1}^{d,\pi ^+}$, $A_{1}^{d,\pi ^-}$, $A_{1}^{p,K ^+}$, $A_{1}^{p,K ^-}$, $A_{1}^{d,K ^+}$ and
$A_{1}^{d,K ^-}$ from HERMES05 \cite{Hermes05}, COMPASS09 \cite{Alekseev:2009ac} and COMPASS10 \cite{Alekseev:2010ub}. The energy scales of SIDIS experimental data are from
$1.16$~GeV${^2}$ to $55.60$~GeV${^2}$.\\
In each collaboration small $z$
data are excluded since the splitting functions in evolution equations lead to negative FFs in their NLO part for
$z\ll 1$, additionally  mass corrections are more important in this region. So we exclude regions where mass corrections and
the singular small $z$ behavior of the splitting functions are effective.
The $z>0.1$ is used for the data which have $\sqrt{s}<M_Z$ and
the $z>0.05$ is used for data which have $\sqrt{s}=M_Z$.
%
\begin{table*}[t]
\caption{\label{tab:lokaonpara}Fit parameters for
the parton FFs into the charged kaon  ($K^+$) at LO ($D_i^{K^+}(z,\mu_0)$) in the  starting scale.}
\begin{ruledtabular}
\begin{tabular}{cccccc}
flavor $i$ &$N_i$ & $\alpha_i$ & $\beta_i$ &$\gamma_i$\\
\hline

$u$ &$ 4.415\pm5.203$&$-0.388\pm0.315$&$ 1.486\pm0.725$&$0.412\pm0.513$
\\
$\overline{s}$ & $29.284\pm15.300$&$1.395\pm1.865$&$ 2.524\pm1.293$&$0.848\pm2.413$
\\
$\overline{u}, d, \overline{d}, s$ & $6.231\pm3.877$&$-0.398\pm0.106$&$ 6.273\pm2.337$&$2.435\pm2.168$
\\
$c, \overline{c}$ &$ 4.853\pm3.343$&$0.245\pm0.411$&$ 4.530\pm1.090$&$-$
\\
$b, \overline{b}$ &$ 8.324\pm2.947$&$0.076\pm0.084$&$8.841\pm1.836$&$-$
\\
$g$              &$ 1.309\pm3.529$&$ 8.871\pm4.530$& $0.293\pm2.44$&$0.165\pm2.732$
\\
\end{tabular}
\end{ruledtabular}
\end{table*}
%
\begin{table*}[t]
\caption{\label{tab:nlokaonpara} Values of fit parameters for the $K^+$ meson at NLO
 in the starting scale.}
\begin{ruledtabular}
\begin{tabular}{cccccc}
flavor $i$ &$N_i$ & $\alpha_i$ & $\beta_i$ &$\gamma_i$\\
\hline
$u$ & $0.660\pm0.156$&$-1.584\pm 0.342$& $0.858\pm0.227$&$0.390\pm0.107$
\\
$\overline{s}$ & $17.769\pm7.775$&$0.708\pm0.390$&$2.479\pm0.316$&$0.665\pm0.218$
\\
$\overline{u}, d, \overline{d}, s$ &$ 6.467\pm1.587$&$0.028\pm0.547$&$ 7.338\pm0.819$&$3.299\pm1.282$
\\
$c, \overline{c}$ & $7.217\pm1.013$&$0.550\pm0.113$&$ 5.366\pm0.314$&$-$
\\
$b, \overline{b}$ &$ 14.675\pm3.227$&$0.293\pm0.080$&$ 10.882\pm0.943$&$-$
\\
$g$              & $2.383\pm0.381$&$5.714\pm0.696$&$ 0.892\pm0.085$& $53542.030\pm5.859$
\\
\end{tabular}
\end{ruledtabular}
\end{table*}
%
\subsection{Neighborhood of global minimum and Hessian method}
\label{subsec:hessian method}
In recent years, the assessment of uncertainties is significant progress in QCD analysis of PDFs and FFs
~\cite{Martin:2009iq,Pumplin:2000vx,Epele:2012vg} and among different approaches Lagrange Multiplier
(LM) technique and Hessian method are the most reliable ones.
While LM technique avoids any approximations or assumptions about the behavior of the $\chi^{2}$ on the
parameters, the only drawback to this method is that its calculation is slow because it needs a separate
minimizations for each parameters.\\
Since we use Hessian or error matrix approach in our analysis, the outline of this method is
explained. The basic assumption of Hessian approach is a quadratic expansion of the $\chi^{2}$ in
the fit parameters ${a_{i}}$ near the global minimum
\begin{equation} \label{eq:hessian}
  \Delta\chi^2 \equiv \chi^2 - \chi_{\rm min}^2 = \sum_{i,j=1}^n H_{ij}(a_i-a_i^0)(a_j-a_j^0),
\end{equation}
with
\begin{equation} \label{eq:hessianmatrix}
  H_{ij} = \left.\frac{1}{2}\frac{\partial^2\,\chi^2}{\partial a_i\partial a_j}\right|_{\rm min},
\end{equation}
where $H_{ij}$ are the elements of the Hessian matrix. Since Hessian matrix and its inverse ($C\equiv H^{-1}$ ), which is the error
matrix, are symmetric, they have a set of $n$ {\it orthogonal} eigenvectors $v_{ik}$ with eigenvalues $\lambda_{k}$
\begin{eqnarray}
\label{eq:eigeq}
\sum_{j=1}^n C_{ij} v_{jk} &=& \lambda_k v_{ik},\\
\label{eq:eivec}
\sum_{i=1}^n v_{ij} v_{ik} &=& \delta_{jk}.
\end{eqnarray}
The parameter variation around the global minimum can be expanded in a basis of eigenvectors and
eigenvalues, that is,
\begin{equation} \label{eq:component}
a_i - a_i^0 = \sum_{k=1}^n e_{ik} z_k,
\end{equation}
where $e_{ik}\equiv \sqrt{\lambda_k}v_{ik}$. Using Eqs.~\ref{eq:eivec} and \ref{eq:component} it can be shown that
the expansion of the $\chi^{2}$ in
the fit parameters ${a_{i}}$ near the global minimum Eq.~\ref{eq:hessian} reduces to
\begin{equation} \label{eq:hessiandiag}
  \chi^2 = \chi^2_{\rm min} + \sum_{k=1}^n z_k^2,
\end{equation}
 where $\sum_{k=1}^n z_k^2\le T^2$ is the interior of a sphere of radius $T$.
 The eigenvector sets $S_k^\pm$ are defined by choosing $T = (\Delta\chi^2)^{1/2}$
 and corresponding positive and negative of eigenvector directions are defined as following
\begin{equation} \label{eq:ziand delta}
  z_i(S_k^\pm) = \pm T\delta_{ik}.
\end{equation}
Using the last equation, the ${a_i}$ parameters that specify the eigenvector basis sets $S_k^\pm$
at a fixed value of $\alpha_S$, are given by
\begin{equation} \label{eq:eigenstept}
  a_i(S_k^\pm) = a_i^0 \pm T\,e_{ik}.
\end{equation}
In the standard parameter-fitting criterion, the errors are given
by the choice of tolerance $T = \Delta \chi ^2 = 1$. Also we can determine the size of uncertainties
applying Hessian method based on correspondence between the confidence
level $P$ and $\Delta \chi ^2$ with the number of fitting parameters $N$
\begin{equation}
        P=\int_0^{\Delta \chi^2} \frac{1}{2\ \Gamma(N/2)}
        \left(\frac{x}{2}\right)^{\frac{N}{2}-1}
               \exp\left(-\frac{x}{2} \right) dx,
\label{eq:dchi2}
\end{equation}
where we get $P = 0.68$ as the confidence level and $\Delta \chi ^2=22.43$ and $\Delta \chi ^2=24.58$ are obtained
for pion and kaon, respectively.\\
The uncertainty on a quantity $F({a_i})$ which is attributive function of the input parameters obtained in the QCD
fit procedure at the scale $Q_0$, is obtained applying the simple Hessian method
\begin{equation}\label{ferror}
\Delta F =T\sqrt{\sum ^{n}_{i,j=1}\frac{\partial F}{\partial a_i}C_{ij}\frac{\partial F}{\partial a_j}}.
\end{equation}
\subsection{The method of error calculation}
\label{subsec:errorcalculation}
According to Eqs.~(\ref{pifit1}-\ref{kafit2}) the evolved fragmentation functions for pion and kaon
are attributive functions of the input parameters which are calculated from the fit.
Their standard linear errors are given by Gaussian error propagation. If $D_{i}^{H}(z;Q^2)$
is the evolved fragmentation density at $Q^2$ then Gaussian error propagation is defined as
\begin{equation} \label{eq:heserror}
[\delta D_{i}^{H}(z)]^2 =\Delta \chi^2{\sum_{j,k}^{n}\frac{\partial D_{i}^{H}(z,a_j)}{\partial a_j}(H_{jk})^{-1}\frac{\partial  D_{i}^{H}(z,a_k)}{\partial a_k}},
\end{equation}
where $\Delta \chi^2$ is the allowed variation in $\chi^2$ and $a_j\mid^{n}_{j=1}$ are free parameters and $n$ is the number of parameters in the global fit. Also
$H_{jk}$ is Hessian or covariance matrix of the parameters determined in the QCD analysis at the initial scale $Q^{2}_{0}$
and it is defined in Eq.~\ref{eq:hessianmatrix}.\\
Consequently we can calculate the uncertainties of any FFs by using Hessian or covariance matrix based on the Gaussian method
at any value of $Q^{2}$ by the QCD evolution. More information and detailed discussion can be found in Refs.\cite{Pumplin:2001ct,Hirai:2003pm,Hirai:2007cx}.
%
\begin{table}[t!]
\caption{\label{tab:exppionLO}The individual $\chi^2$ values and the fitted normalization in the LO for
each collaboration and the total $\chi^2$ fit for $\pi^+$.}
\begin{ruledtabular}
\tabcolsep=0.06cm \footnotesize
\begin{tabular}{lccccc}
Collaboration& data & $\sqrt{s}$  GeV&  data   & Relative  & $\chi^2$(LO)\\
             & properties &          &  points & normalization &        \\
             &    &                  &         & in fit &              \\\hline
Belle \cite{Leitgab:2013qh}  &   untagged   & 10.52      & 78 & 0.983 & 11.5  \\
BaBar \cite{Lees:2013rqd}  &   untagged   & 10.54      & 38 & 0.936 & 204.5  \\
TPC \cite{tpc29}  &   untagged   & 29      & 12 &  0.993 & 6.4  \\
TASSO \cite{tasso34_44}  & untagged \  &  34 & 8 & 1.051 & 6.6\\
          & untagged         & 44 &  5 & 1.051 &6.2\\
TOPAZ \cite{topaz58}  &   untagged   & 58      & 4 &  1.013&1.3\\
ALEPH \cite{aleph91}    & untagged  & 91.2 & 22 &  0.997 &28.2 \\
OPAL \cite{opal91}  & untagged\ & 91.2 & 22 & 1.017&35.3\\
SLD \cite{sld91}  & untagged\  & 91.2 & 29 & 1.012 &53.6\\
          & $uds$ tagged         &  91.2 & 29 & 1.012&94.7 \\
          & $c$ tagged           &  91.2 & 29 & 1.012&44.8\\
          & $b$ tagged           &  91.2 & 29 & 1.012 &90.6\\
DELPHI \cite{delphi91,delphi91-2}  & untagged & 91.2  & 17 & 0.987&9.4 \\
          & $uds$ tagged         &  91.2 & 17 & 0.987&7.7 \\
          & $b$ tagged           &  91.2 & 17 & 0.987 &53.7\\
HERMES \cite{Hermes05}  & SIDIS(p,$\pi^{+}$) & 1.10-3.23  & 9 & 1.051&10.1 \\
          & SIDIS(p,$\pi^{-}$)       &  1.10-3.23 & 9 & 1.051 &6.6\\
          & SIDIS(d,$\pi^{+}$)       &  1.10-3.2 & 9 & 1.051&11.4\\
          &SIDIS(d,$\pi^{-}$)       &  1.10-3.2 & 9 & 1.051 &20.5\\
COMPASS \cite{Alekseev:2009ac}  & SIDIS(d,$\pi^{+}$) & 1.07-5.72  & 10 & 1.028&3.9 \\
          & SIDIS(d,$\pi^{-}$) & 1.07-5.72  & 10 & 1.028&5.5 \\
COMPASS \cite{Alekseev:2010ub}  & SIDIS(p,$\pi^{+}$) & 1.07-7.45  & 12 & 0.997&10.8 \\
          & SIDIS(p,$\pi^{-}$) & 1.07-7.45  & 12 & 0.997&13.7 \\
{\bf TOTAL:} & & &436 &     &736.32\\
($\chi^{2}$/ d.o.f ) & & & & &1.77\\
\end{tabular}
\end{ruledtabular}
\end{table}
%
%
\begin{table}[b!]
\caption{\label{tab:exppionNLO}The individual $\chi^2$ values and the fitted normalization in the NLO for
each collaboration and the total $\chi^2$ fit for $\pi^+$.}
\begin{ruledtabular}
\tabcolsep=0.06cm \footnotesize
\begin{tabular}{lccccc}
Collaboration& data & $\sqrt{s}$  GeV&  data   & Relative  & $\chi^2$(NLO)\\
             & properties &          &  points & normalization &        \\
             &    &                  &         & in fit &              \\\hline
Belle \cite{Leitgab:2013qh}  &   untagged   & 10.52      & 78 & 1.001 & 12.5  \\
BaBar \cite{Lees:2013rqd}  &   untagged   & 10.54      & 38 & 0.928 & 138.3  \\
TPC \cite{tpc29}  &   untagged   & 29      & 12 &  0.992 & 5.7  \\
TASSO \cite{tasso34_44}  & untagged \  &  34 & 8 & 1.049 & 7.9\\
          & untagged         & 44 &  5 & 1.049 &6.9\\
TOPAZ \cite{topaz58}  &   untagged   & 58      & 4 & 1.015& 1.6\\
ALEPH \cite{aleph91}    & untagged  & 91.2 & 22 &  1.001 &31.7 \\
OPAL \cite{opal91}  & untagged\ & 91.2 & 22 & 1.020& 33.5\\
SLD \cite{sld91}  & untagged\  & 91.2 & 29 & 1.015 &31.7\\
          & $uds$ tagged         &  91.2 & 29 & 1.015&62.3 \\
          & $c$ tagged           &  91.2 & 29 &1.015&26.8\\
          & $b$ tagged           &  91.2 & 29 & 1.015 & 85.2\\
DELPHI \cite{delphi91,delphi91-2}  & untagged & 91.2  & 17 & 0.991&15.9 \\
          & $uds$ tagged         &  91.2 & 17 & 0.991&13.2 \\
          & $b$ tagged           &  91.2 & 17 & 0.991 &48.8\\
HERMES \cite{Hermes05}  & SIDIS(p,$\pi^{+}$) & 1.10-3.23  & 9 & 1.063&10.3 \\
          & SIDIS(p,$\pi^{-}$)       &  1.10-3.23 & 9 & 1.063 &4.6\\
          & SIDIS(d,$\pi^{+}$)       &  1.10-3.2 & 9 & 1.063&18.6 \\
          &SIDIS(d,$\pi^{-}$)       &  1.10-3.2 & 9 & 1.063 &22.3\\
COMPASS \cite{Alekseev:2009ac}  & SIDIS(d,$\pi^{+}$) & 1.07-5.72  & 10 & 1.071&12.8 \\
          & SIDIS(d,$\pi^{-}$) & 1.07-5.72  & 10 & 1.071&5.61 \\
COMPASS \cite{Alekseev:2010ub}  & SIDIS(p,$\pi^{+}$) & 1.07-7.45  & 12 & 1.011&10.5 \\
          & SIDIS(p,$\pi^{-}$) & 1.07-7.45  & 12 & 1.011&7.6 \\
{\bf TOTAL:} & & &436 &     &611.52\\
($\chi^{2}$/ d.o.f ) & & & & &1.47\\
\end{tabular}
\end{ruledtabular}
\end{table}
\section{Inclusive  $\pi^+/K^+$-Mesons production in Top-quark decay}
\label{sec:four}
Nowadays, the CERN Large Hadron Collider (LHC) is a superlative machine
to produce top-quark pairs so that  at design energy $\sqrt{S}=14$~TeV and design luminosity
${\cal L}=10^{34}$~cm$^{-2}$s$^{-1}$ in each of the four experiments
it is expected to produce about 90 million top-quark pairs per year \cite{Moch:2008qy}.
This much volume of events will allow us to determine the properties of the top-quark, such as its
mass $m_t$, branching fractions and matrix elements $V_{tq}$
 of the Cabibbo-Kobayashi-Maskawa (CKM) \cite{Cabibbo:1963yz} with high precision.
Because of its large mass, the top-quark decays so rapidly so that it has no time to
hadronize and due to $|V_{tb}|\approx1$, top-quarks almost  decay to bottom-
quarks, via the decay mode $t\to bW^+$ in the standard model (SM). Bottom-quarks also hadronize,
via $b\to H+X$, before they decay,
and thus the decay mode $t\to HW^++X$ is of prime importance where H refers to the detected outgoing
hadron. A particular interest at LHC is to study the scaled energy distributions of outgoing hadron.\\
In this section, we study the energy spectrum of the inclusive light mesons including
$\pi^+$ and $K^+$ in  top-quark decay working  in the zero-mass variable-flavor-number
scheme(ZM-VFNs). \\
We wish to study the inclusive production of a light meson in the decay process
\begin{eqnarray}\label{pros}
t\rightarrow b+W^+ (g)\rightarrow \pi^\pm/K^\pm+X,
\end{eqnarray}
where $X$ stands for the unobserved final state. The gluon in Eq.~(\ref{pros}) contributes
 to the real radiation at NLO and both the $b$-quark and the gluon may hadronize to the outgoing light mesons.
 To obtain the energy distribution of light hadrons we use the realistic FFs
obtained in our approach. \\
To study the energy spectrum  of outgoing meson it would be convenient to introduce the
 scaled energy fractions  $x_i=E_i/E_b^\text{max}$ ($i=b,g,H$)
where $H$ stands for the light mesons. In the top-qurak rest frame, the energies range
$0\le (E_b,E_g)\le (m_t^2-m^2_W)/(2m_t) $  and
$m_H\le E_H\le (m_t^2+m_H^2-m^2_{W})/(2m_t) $ \cite{Kniehl:2012mn}.\\
We wish to calculate the partial decay width of process~(\ref{pros}) differential in
$x_H$ ($d\Gamma/dx_H$) at NLO in the ZM-VFN scheme.
Considering the factorization  theorem of the QCD \cite{Collins:1998rz}, the energy distribution of a hadron H can
be expressed as the convolution of the parton-level spectrum with the
 fragmentation densities $D_i^H(z,\mu_F)$,  describing the hadronization $i\rightarrow H$,
\begin{equation}
\frac{d\Gamma}{dx_H}=\sum_{i=b,g}\int_{x_i^\text{min}}^{x_i^\text{max}}
\frac{dx_i}{x_i}\,\frac{d\hat\Gamma_i}{dx_i}(\mu_R,\mu_F)
D_i^H\left(\frac{x_H}{x_i},\mu_F\right),
\label{eq:master}
\end{equation}
where $d\hat\Gamma_i/dx_i $ is the parton-level differential width  of the process
$t\to i+W^+ (i=b,g)$, which are being extracted from Ref.~\cite{Kniehl:2012mn}.
Here, $\mu_F$ and $\mu_R$ are the factorization and the renormalization scales
which are set to $\mu_R=\mu_F=m_t$.
The values of all FF parameters are listed in
Tables~\ref{tab:lopionpara}, \ref{tab:nlopionpara}, \ref{tab:lokaonpara} and \ref{tab:nlokaonpara}.
Since  these FFs are parameterized at the low factorization scale, extraction of the FFs at each
arbitrary scale of energy should be performed using the grids and
FORTRAN routines based on solving DGLAP equations.
\section{Fit results}
\label{sec:five}
Now we are in a situation to explain our global analysis of fragmentation functions for pions and kaons results. We compare our calculated cross
section and double spin asymmetry results with the experimental data and find a good agreement.

%
\begin{table}[t!]
\caption{\label{tab:expkaonLO} The individual $\chi^2$ values and the fitted normalization in the LO for
each collaboration and the total $\chi^2$ fit for $K^+$.}
\begin{ruledtabular}
\tabcolsep=0.06cm \footnotesize
\begin{tabular}{lccccc}
Collaboration& data & $\sqrt{s}$  GeV&  data   & Relative  & $\chi^2$(LO)\\
             & properties &          &  points & normalization &        \\
             &    &                  &         & in fit &              \\\hline
Belle \cite{Leitgab:2013qh}  &   untagged   & 10.52      & 78 & 1.060 & 109.1  \\
BaBar \cite{Lees:2013rqd}  &   untagged   & 10.54      & 28 & 0.992 & 130.5  \\
TPC \cite{tpc29}  &   untagged   & 29      & 11 &  1.063&9.0\\
TASSO \cite{tasso34_44}  & untagged \  &  34 & 4 & 0.995& 0.3\\
TOPAZ \cite{topaz58}  &   untagged   & 58      & 3 & 1.004&0.8 \\
ALEPH \cite{aleph91}    & untagged  & 91.2 & 18 &   1.008&6.1\\
OPAL \cite{opal91}  & untagged\ & 91.2 & 10 & 0.975&2.7\\
SLD \cite{sld91}  & untagged\  & 91.2 & 29 & 0.993&19.3\\
          & $uds$ tagged         &  91.2 & 29 & 0.993&61.1\\
          & $c$ tagged           &  91.2 & 29 & 0.993&37.2\\
          & $b$ tagged           &  91.2 & 28 &0.993&153.6\\
DELPHI \cite{delphi91,delphi91-2}  & untagged & 91.2  & 17 &1.063& 2.3\\
          & $uds$ tagged         &  91.2 & 17 &  1.063&7.4\\
          & $b$ tagged           &  91.2 & 17 & 1.063&11.9\\
HERMES \cite{Hermes05}  & SIDIS(d,$K^{+}$) & 1.22-3.19  & 9 & 1.004&7.1 \\
          & SIDIS(d,$K^{-}$)       &  1.22-3.19 & 9 & 1.004 &6.8\\
COMPASS \cite{Alekseev:2009ac}  & SIDIS(d,$K^{+}$) & 1.07-5.72  & 10 & 1.012&7.9 \\
          & SIDIS(d,$K^{-}$) & 1.07-5.72  & 10 & 1.012&16.0 \\
COMPASS \cite{Alekseev:2010ub}  & SIDIS(p,$K^{+}$) & 1.07-7.45  & 12 & 1.004&10.8 \\
          & SIDIS(p,$K^{-}$) & 1.07-7.45  & 12 & 1.004& 12.26 \\
{\bf TOTAL:} & & &380 &     &612.18\\
($\chi^{2}$/ d.o.f ) & & & & &1.71\\
\end{tabular}
\end{ruledtabular}
\end{table}
%
%
%
\begin{table}[b!]
\hspace{-0.2cm}
\caption{\label{tab:expkaonNLO} The individual $\chi^2$ values and the fitted normalization in the NLO for
each collaboration and the total $\chi^2$ fit for $K^+$.}
\begin{ruledtabular}
\tabcolsep=0.05cm \footnotesize
\begin{tabular}{lccccc}
Collaboration& data & $\sqrt{s}$  GeV&  data   & Relative  & $\chi^2$(NLO)\\
             & properties &          &  points & normalization &        \\
             &    &                  &         & in fit &              \\\hline
Belle \cite{Leitgab:2013qh}  &   untagged   & 10.52      & 78 & 1.029 & 105.7  \\
BaBar \cite{Lees:2013rqd}  &   untagged   & 10.54      & 28 & 0.974 & 100.8 \\
TPC \cite{tpc29}  &   untagged   & 29      & 11 &  1.041& 7.6\\
TASSO \cite{tasso34_44}  & untagged \  &  34 & 4 & 0.992& 0.3\\
TOPAZ \cite{topaz58}  &   untagged   & 58      & 3 & 1.004&0.8 \\
ALEPH \cite{aleph91}    & untagged  & 91.2 & 18 &   1.017&4.9\\
OPAL \cite{opal91}  & untagged\ & 91.2 & 10 & 0.983&2.8\\
SLD \cite{sld91}  & untagged\  & 91.2 & 29 & 1.003&19.1\\
          & $uds$ tagged         &  91.2 & 29 & 1.003&67.6\\
          & $c$ tagged           &  91.2 & 29 & 1.003&42.7\\
          & $b$ tagged           &  91.2 & 28 & 1.003&116.5\\
DELPHI \cite{delphi91,delphi91-2}  & untagged & 91.2  & 17 &1.084& 2.6\\
          & $uds$ tagged         &  91.2 & 17 &  1.084&6.9\\
          & $b$ tagged           &  91.2 & 17 & 1.084&11.8\\
HERMES \cite{Hermes05}  & SIDIS(d,$K^{+}$) & 1.22-3.19  & 9 & 1.009&6.0 \\
          & SIDIS(d,$K^{-}$)       &  1.22-3.19 & 9 & 1.009 &8.1\\
COMPASS \cite{Alekseev:2009ac}  & SIDIS(d,$K^{+}$) & 1.07-5.72  & 10 & 1.032&7.2 \\
          & SIDIS(d,$K^{-}$) & 1.07-5.72  & 10 & 1.032&20.4 \\
COMPASS \cite{Alekseev:2010ub}  & SIDIS(p,$K^{+}$) & 1.07-7.45  & 12 & 1.013&6.5 \\
          & SIDIS(p,$K^{-}$) & 1.07-7.45  & 12 & 1.013&15.3 \\
{\bf TOTAL:} & & &380 &     &551.32\\
($\chi^{2}$/ d.o.f ) & & & & &1.54\\
\end{tabular}
\end{ruledtabular}
\end{table}
According to
last section that we introduce the experimental data for our fits, we present
our results of the optimum fits for the fragmentation parameters of the $\pi^+$- and $K^+$-mesons
in the initial scale $\mu_0$ at LO and NLO.
Also $\chi^2$ and normalization fit values for each individual collaboration are reported.
On the other hand we are interested in presenting how much the asymmetry SIDIS data effect for determination of
pion and kaon FFs. Also the FFs comparisons are made with the results obtained in the other FFs analysis in Refs.~\cite{Hirai:2007cx,deFlorian:2007aj,Albino:2008fy}. We also briefly present the dependance
of $\Delta\chi^{2}$ global along some random samples of eigenvector
directions to illustrate the deviations of the $\Delta\chi^{2}$ function from the expected quadratic dependence.
At the end we show our prediction for energy spectrum of pion and kaon as light mesons in top decay.
\begin{figure*}[h!]
\vspace{-0.4cm}
\centerline{\includegraphics[width=0.65\textwidth,angle=-90]{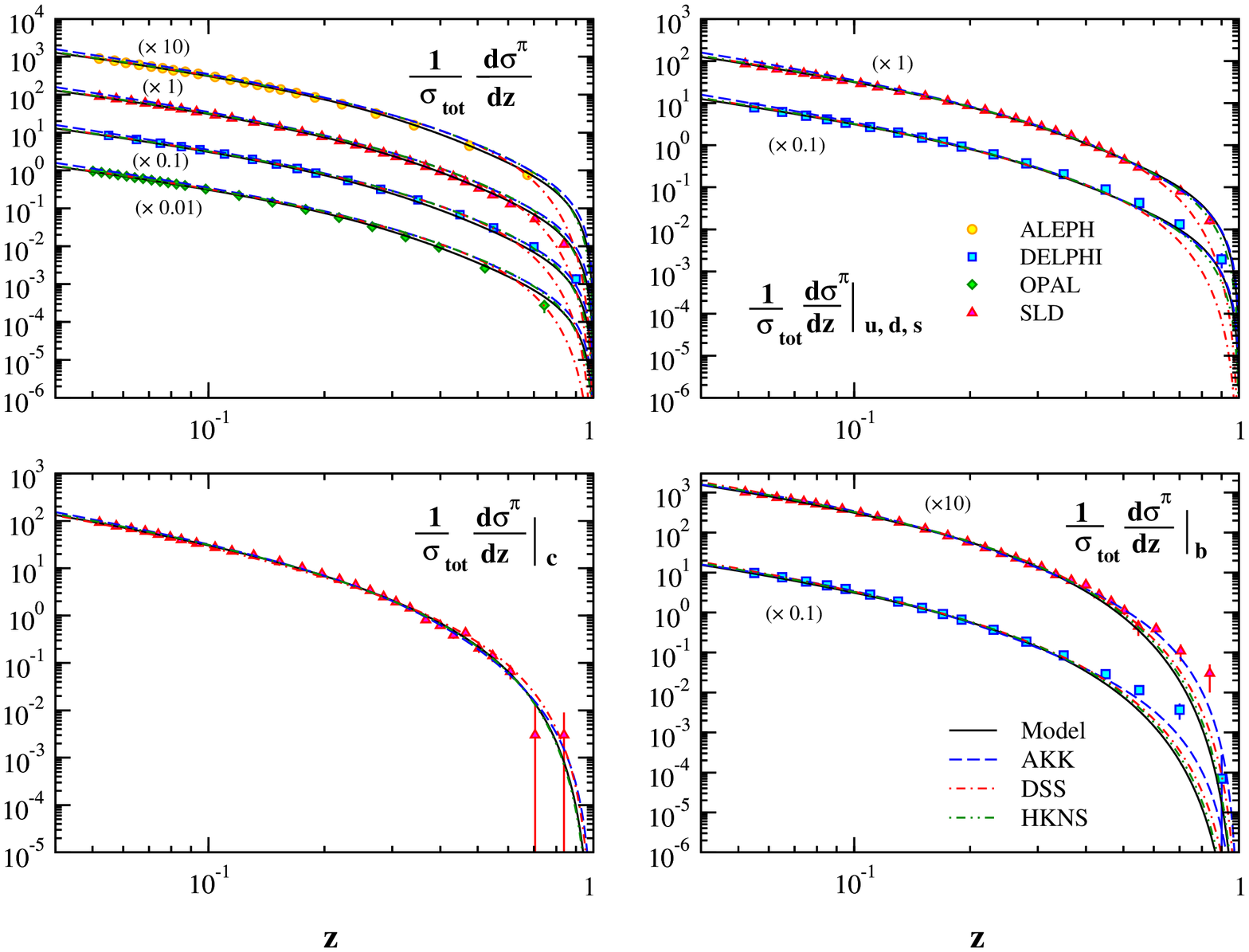}}
\vspace*{-0.7cm}
\caption{Comparison of our NLO results for $\frac{1}{\sigma_{tot}}\frac{d\sigma^{\pi}}{dz}$ in
total and tagged cross sections with pion production data at $Q^2=M_{Z}^{2}$ by ALEPH, DELPHI, OPAL and SLD~\cite{aleph91,delphi91,delphi91-2,opal91,sld91}. Our model, is denoted Model, also is compered with the other models~\cite{Albino:2008fy,deFlorian:2007aj,Hirai:2007cx}.}
\label{pion+data}
\vspace*{-0.3cm}
\centerline{\includegraphics[width=0.65\textwidth,angle=-90]{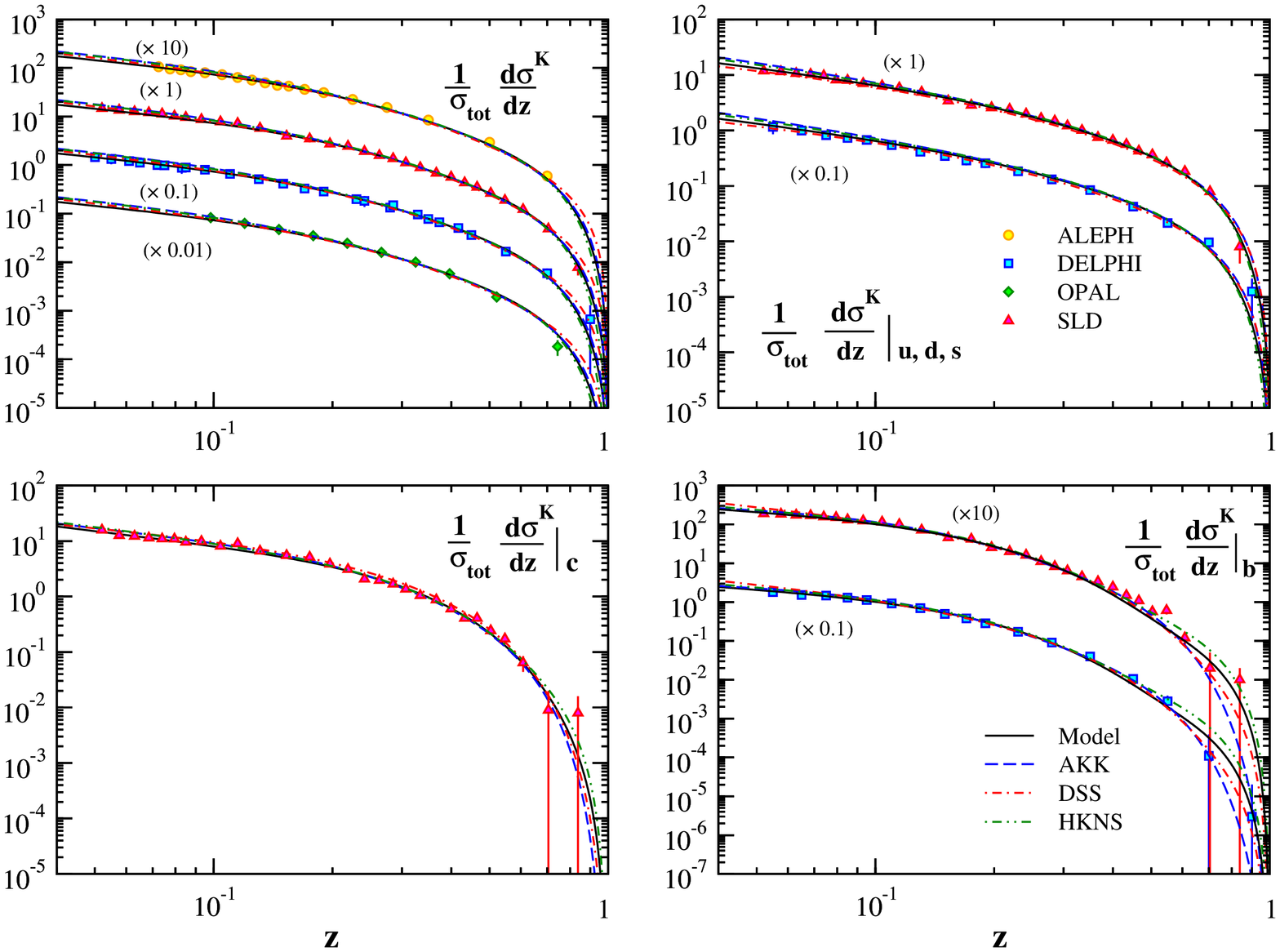}}
\vspace*{-0.7cm}
\caption{ Comparison of our NLO results for $\frac{1}{\sigma_{tot}}\frac{d\sigma^{K}}{dz}$ in
total and tagged cross sections with kaon production data at $Q^2=M_{Z}^{2}$ by ALEPH, DELPHI, OPAL and SLD~\cite{aleph91,delphi91,delphi91-2,opal91,sld91}. Our model, is denoted Model, also is compered with the other models~\cite{Albino:2008fy,deFlorian:2007aj,Hirai:2007cx}.}
\label{kaon+data}
\vspace*{-0.5cm}
\end{figure*}
\begin{figure*}[t!]
\vspace{-2cm}
\centerline{\includegraphics[width=0.7\textwidth,angle=-90]{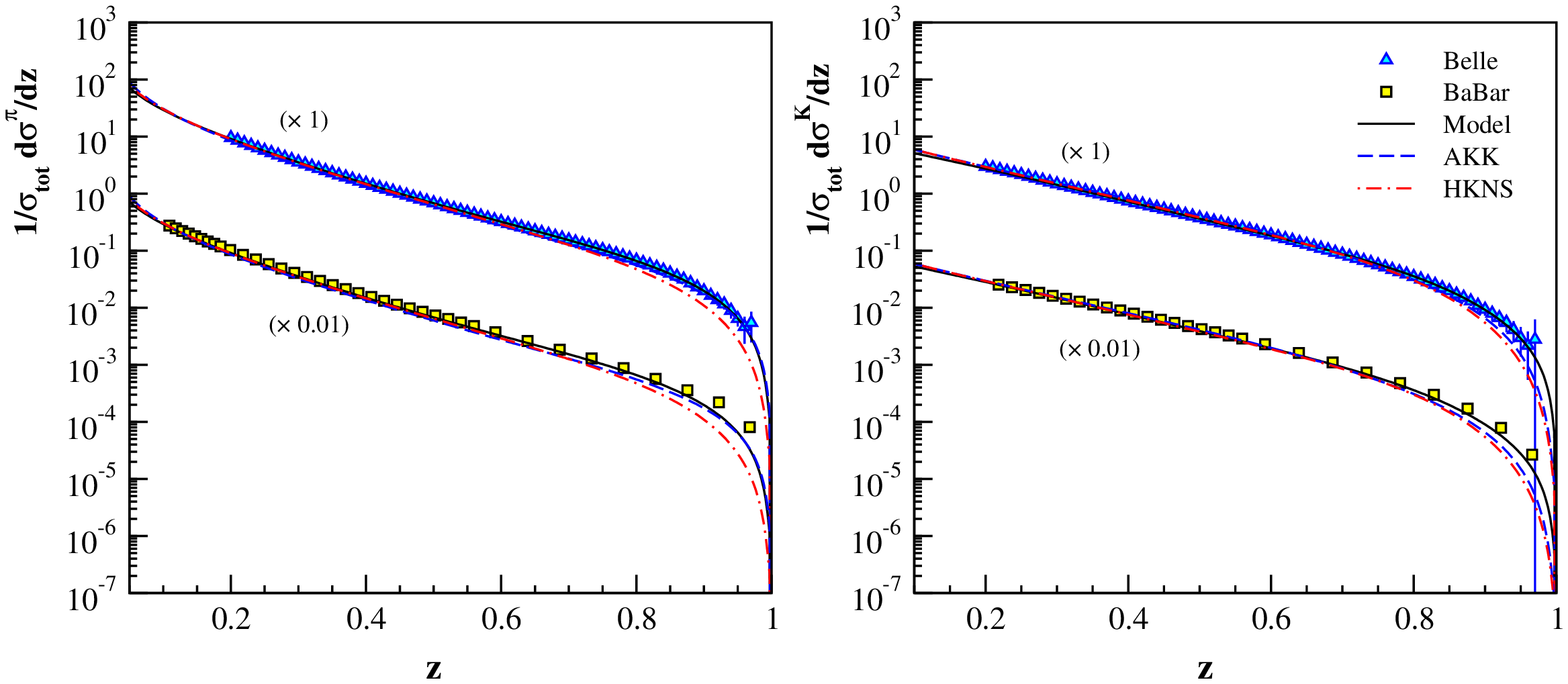}}
\vspace{-3cm}
\caption{Comparison of our NLO results for $\frac{1}{\sigma_{tot}}\frac{d\sigma^{i}}{dz},~ (i=\pi, K)$ in
total cross sections with pion and kaon production data at $Q=10.54$~GeV by BaBar~\cite{Lees:2013rqd} and $Q=10.52$~GeV by Belle~\cite{Leitgab:2013qh}. Our model, is denoted Model, also is compered with the other models~\cite{Albino:2008fy,Hirai:2007cx}.}
\label{BBdata}
\end{figure*}
%
\subsection{Experimental data and our analysis}
\label{subsec:data}
\begin{figure}[t!]
\centerline{\includegraphics[width=0.55\textwidth,angle=-90]{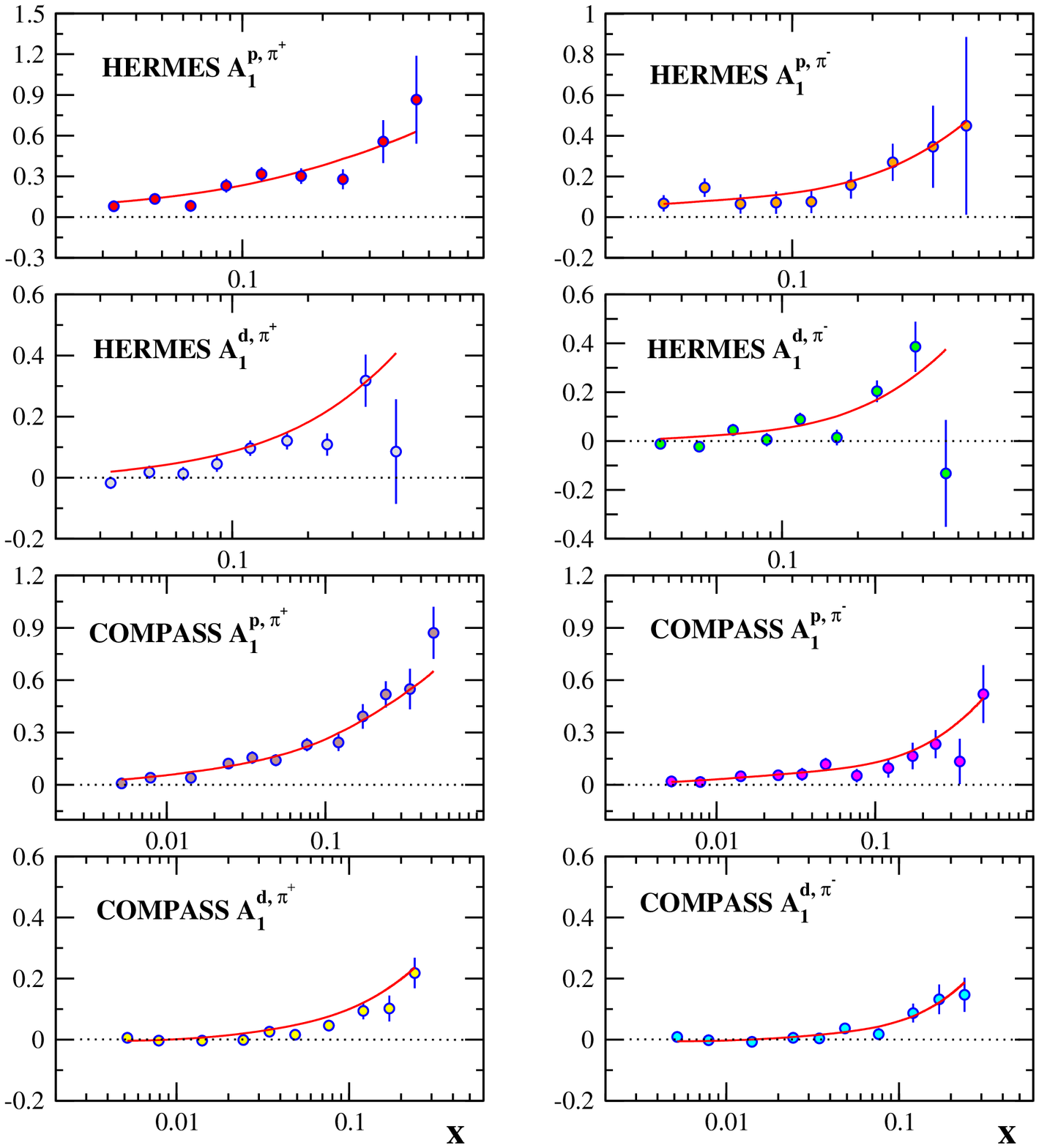}}
\vspace*{+0.1cm}
\caption{Double spin asymmetry SIDIS data for $\pi^+$
and $\pi^-$ from HERMES and COMPASS~\cite{Hermes05,Alekseev:2010ub,Alekseev:2009ac} at measured $x$ and $Q^2$ and comparison with the fit results
of our global analysis at NLO.}
\label{pi}
\vspace*{-0.1cm}
\centerline{\includegraphics[width=0.49\textwidth,angle=-90]{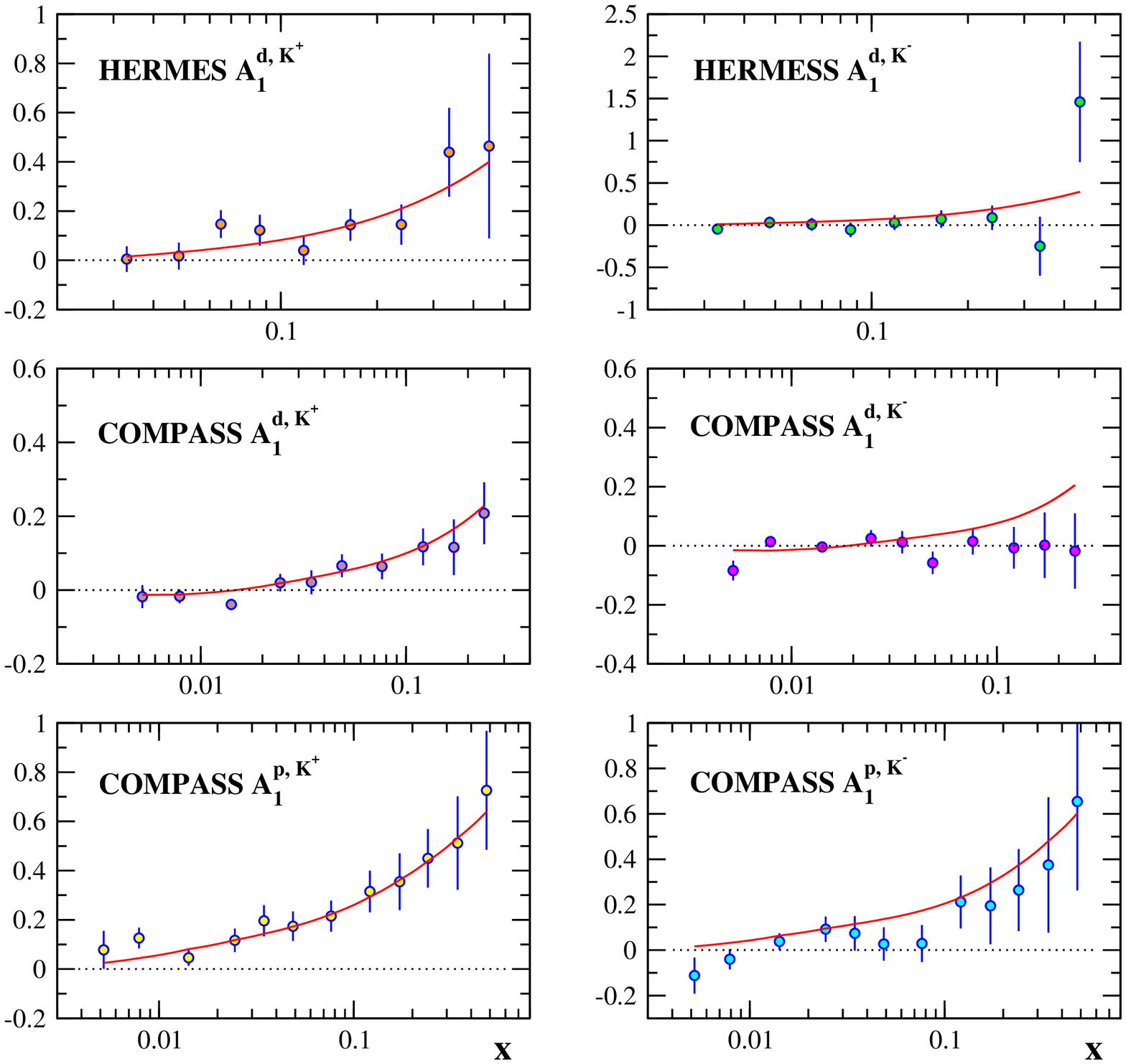}}
\vspace*{-0.1cm}
\caption{Double spin asymmetry SIDIS data for $K^+$ and
$K^-$ from HERMES and COMPASS~\cite{Hermes05,Alekseev:2010ub,Alekseev:2009ac} at measured $x$ and $Q^2$ and comparison with the fit results of our
global analysis at NLO.}
\label{ki}
\end{figure}

Experimental data for inclusive hadron production
in SIA cover orders of energy magnitude from $10.52$ GeV to the mass of the $Z$-boson. By adding SIDIS experimental data
the range of energy is extended and cover low energy from $1.16$~GeV${^2}$ to $55.60$~GeV${^2}$.
In Figs.~\ref{pion+data} and \ref{kaon+data}, using new  functional form of FFs we compare our results for $\frac{1}{\sigma_{tot}}\frac{d\sigma^{i}}{dz}$ ($i=\pi$ or $K$) with the data at $\mu^2=M_Z^{2}$ reported by ALEPH, DELPHI, OPAL and SLD collaborations at NLO.
In these figures we separate the light, charm and bottom tagged cross sections and most of the diagrams show
a remarkable agreement between our results and experimental data.\\
Comparing other FFs models in these figures also gives a nice altogether description of our model with the AKK set~\cite{Albino:2008fy} that
included hadron production data in electron-positron and hadron-hadron scattering data, the DSS set \cite{deFlorian:2007aj} that included electron-positron,
lepton-nucleon and hadron-hadron scattering data and HKNS set~\cite{Hirai:2007cx} that included electron-positron data.
However, as Fig.~\ref{pion+data} shows, due to  the largeness of $\chi^{2}$ contributions for SLD and DELPHI $b$ tagged (see  Table \ref{tab:exppionLO} and \ref{tab:exppionNLO}) some points are outside of the curves. Generally, $\chi^{2}$ values of the heavy flavors, in particular,  $b$ tagged data are larger than the other data (see Tables \ref{tab:exppionLO}, \ref{tab:exppionNLO}, \ref{tab:expkaonLO} and \ref{tab:expkaonNLO}) that it might be caused by some extent to contaminations
from weak decay.\\
Recent differential cross section data from Belle and BaBar collaborations are included in our analysis. In Fig.~\ref{BBdata} our results for pion and kaon at $Q=10.52$~GeV and $Q=10.54$~GeV are compered with these data. Also the other FFs models are compared with Belle and BaBar data and this figure shows a nice agreement between our model and these data.\\
Figs.~\ref{pi} and \ref{ki} present
the extracted values of $A_1$ for proton and deuteron from the global fit for $\pi^+$, $\pi^-$, $K^+$ and $K^-$ at NLO
in comparison with the SIDIS data from
HERMES and COMPASS~\cite{Hermes05,Alekseev:2010ub,Alekseev:2009ac}. The extraction of double spin asymmetry data from the global fit for fragmentation functions is done for the first time and as it is shown the overall agreement of the
experimental data sets in the global analysis is great. Some of the theoretical analysis such as Refs.~\cite{deFlorian:2009vb,lss} use the asymmetry data from DIS and SIDIS to
calculate the polarized parton distributions.
\subsection{Fit results for $\pi^+$ and $K^+$ FFs}
\begin{figure}[t!]
\centerline{\includegraphics[width=0.43\textwidth,angle=-90]{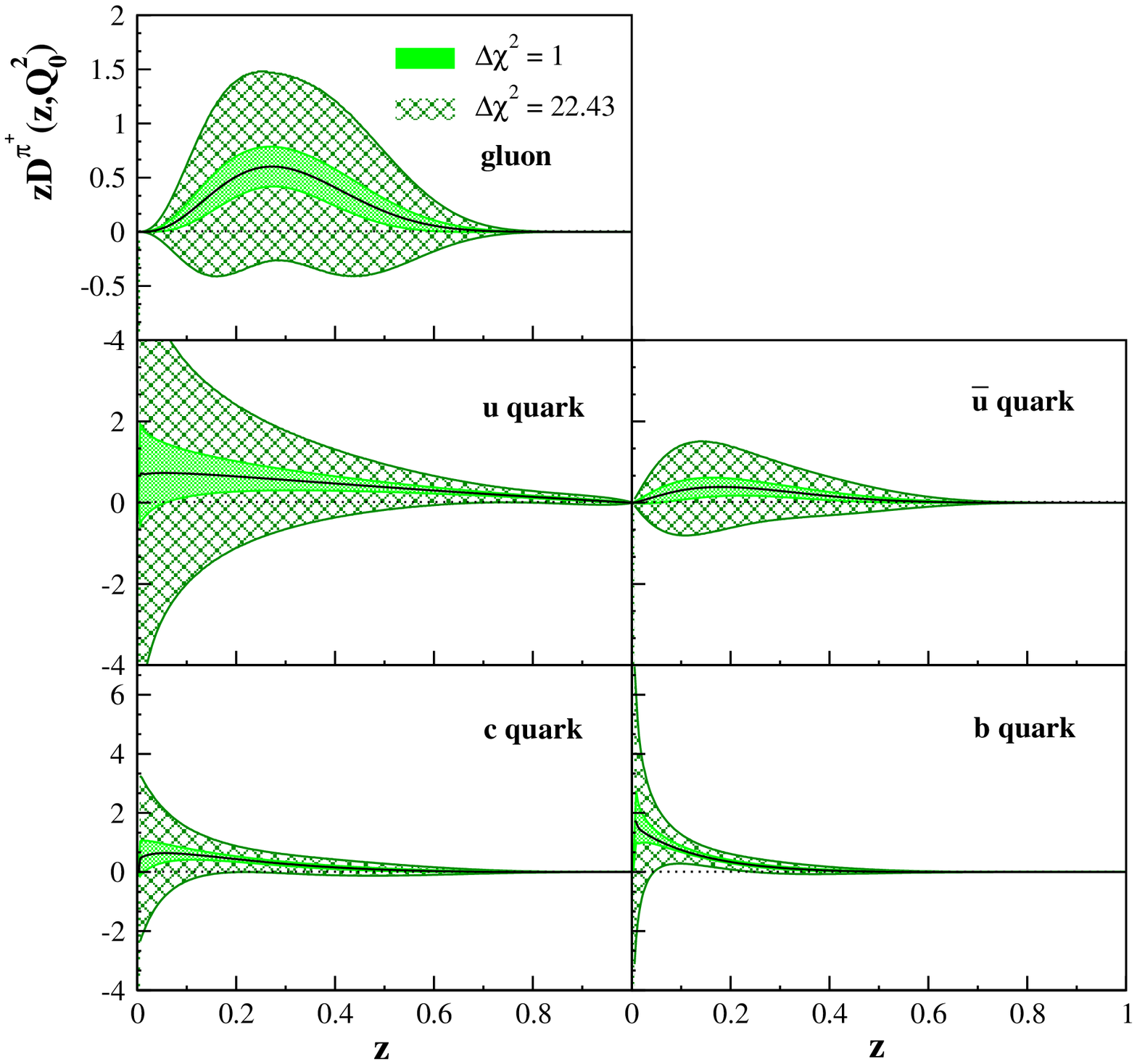}}
\vspace{-0.4cm}
\caption{Fragmentation densities and their uncertainties\\
are shown for $\pi^+$ at
$Q_{0}^{2} = 1$ GeV$^2, m_{c}^{2}$ and $m_{b}^{2}$ at NLO.
Their uncertainties are presented for $\Delta \chi^2 = 1$ (inner
bands) and $\Delta \chi^2 = 22.43$ (outer bands).}
\label{errorpionNLO}
\centerline{\includegraphics[width=0.43\textwidth,angle=-90]{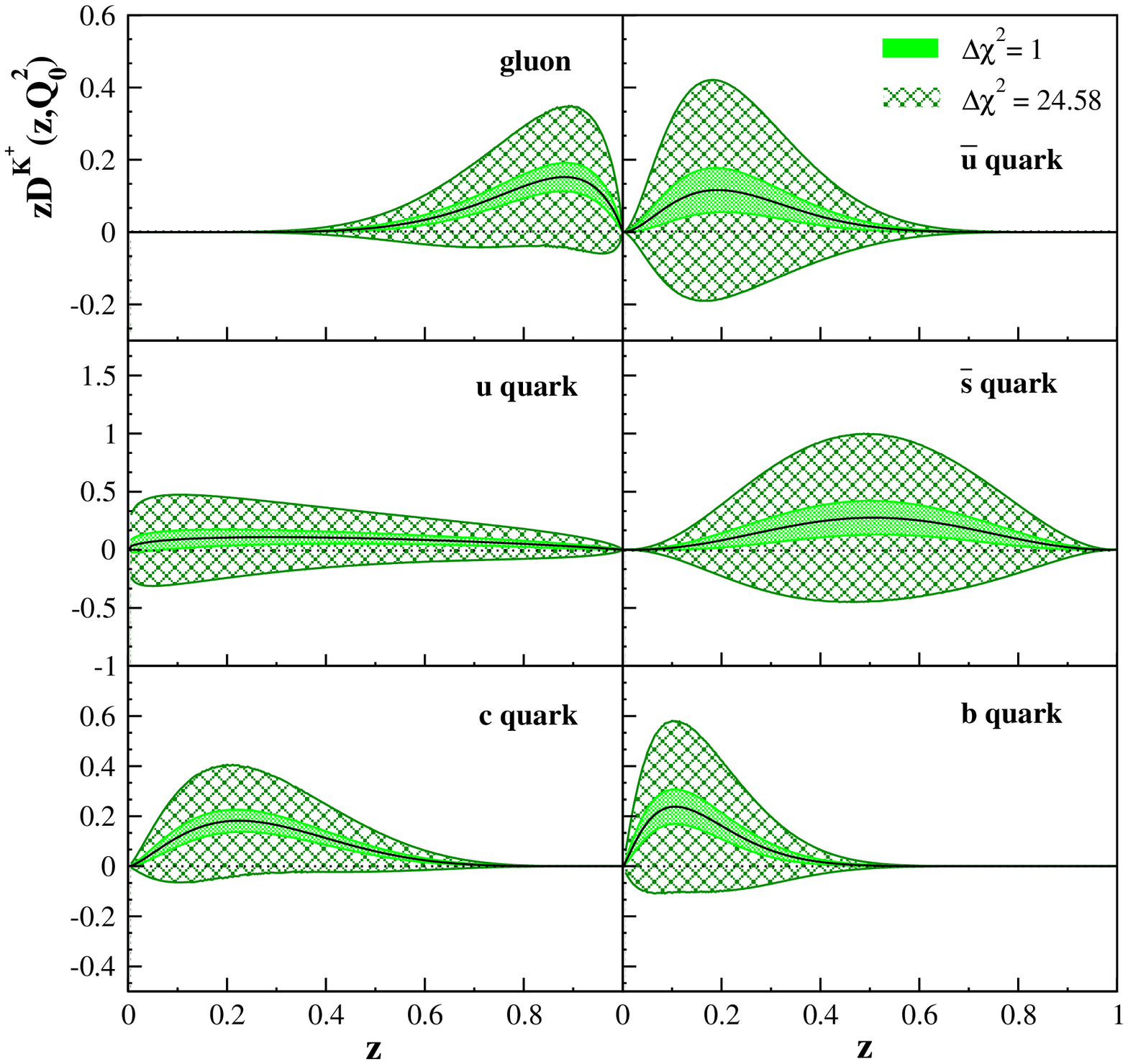}}
\vspace{-0.4cm}
\caption{ Fragmentation densities and their uncertainties\\
are shown for $K^+$ at
$Q_{0}^{2} = 1$ GeV$^2, m_{c}^{2}$ and $m_{b}^{2}$ at NLO.
Their uncertainties are presented for $\Delta \chi^2 = 1$ (inner
bands) and $\Delta \chi^2 = 24.58$ (outer bands).}
\vspace{-0.5cm}
\label{errorkaonNLO}
\end{figure}

According to the scenarios defined for the  fragmentation functions
of  $\pi^+$ and $K^+$-mesons at the starting
scales, 20 and 22 parameters have to be determined, respectively.
These parameters are listed in Tables \ref{tab:lopionpara} and \ref{tab:nlopionpara}
for $\pi^+$ and in Tables \ref{tab:lokaonpara} and \ref{tab:nlokaonpara} for
$K^+$ at LO and NLO. The initial  scales for the $b\rightarrow \pi^+/K^+$ and
$c\rightarrow \pi^+/K^+$ FFs are $\mu_0^{2}=m_b^{2}$ and $\mu_0^{2}=m_c^{2}$, respectively,
and $\mu_{0}^{2}= 1$~GeV$^2$ is considered for the gluon and light-quarks.
In Tables~\ref{tab:exppionLO}, \ref{tab:exppionNLO}, \ref{tab:expkaonLO} and \ref{tab:expkaonNLO}, we list all experimental data
sets included in our global analysis, discussed in Sec.~\ref{subsec:data}, and
the $\chi^{2}$ values per degree of freedom
pertaining to the LO and NLO fits are presented for each collaboration based on data points.
Also the relative normalization calculated of fit
for each data set is reported in these tables. Indeed the global minimization of $\chi^2$, discussed in Sec.~\ref{subsec:chi2}, in the global
fit considerably improves after taking into account relative normalization.\\
The $\pi^+$ and $K^+$ fragmentation densities and their uncertainties are presented in
Figs.~\ref{errorpionNLO} and \ref{errorkaonNLO} at $\mu_{0}^2= 1$~GeV$^2$ for the gluon and light-quarks and $\mu_0^{2}=m_c^{2}$ and $\mu_0^{2}=m_b^{2}$
for $c$ and $b$-quarks at NLO. We present the FFs uncertainties for $\Delta \chi ^2=1$ and $\Delta \chi ^2=22.43$ for pion
and $\Delta \chi ^2=24.58$ for kaon in Figs.~\ref{errorpionNLO} and \ref{errorkaonNLO}. The method of error calculation is described in Secs.~ \ref{subsec:hessian method} and \ref{subsec:errorcalculation}.\\
%
%
%
\begin{figure}[b!]
\vspace{-0.4cm}
\centerline{\includegraphics[width=0.47\textwidth,angle=-90]{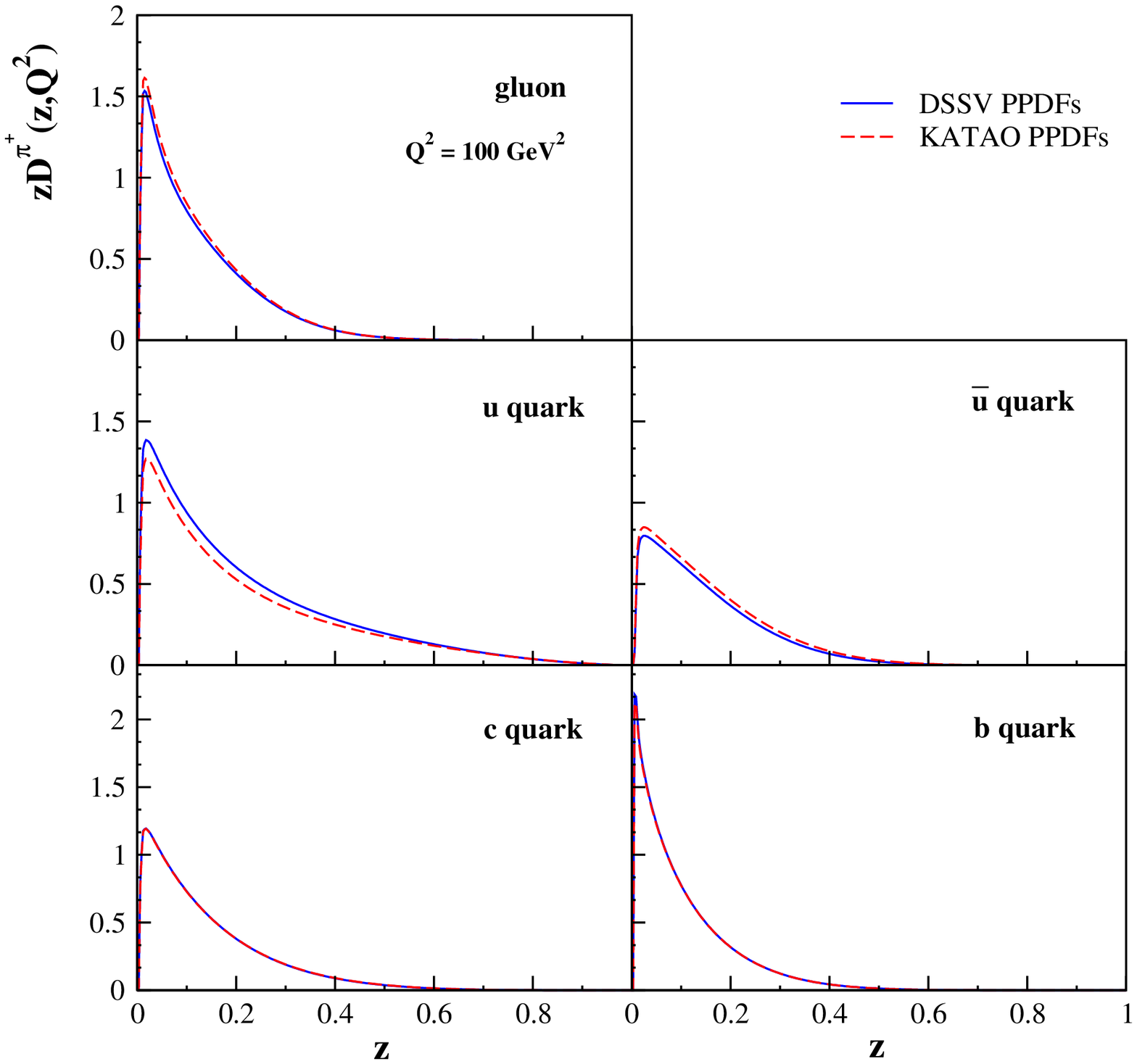}}
\vspace{-0.4cm}
\caption{Comparison of extracted pion fragmentation functions at $Q^{2} = 100$ GeV$^2$ using KATAO \cite{Khorramian:2010qa}
and DSSV \cite{deFlorian:2009vb} PPDFs at NLO.}
\label{pDSSV-KATAO}
\centerline{\includegraphics[width=0.47\textwidth,angle=-90]{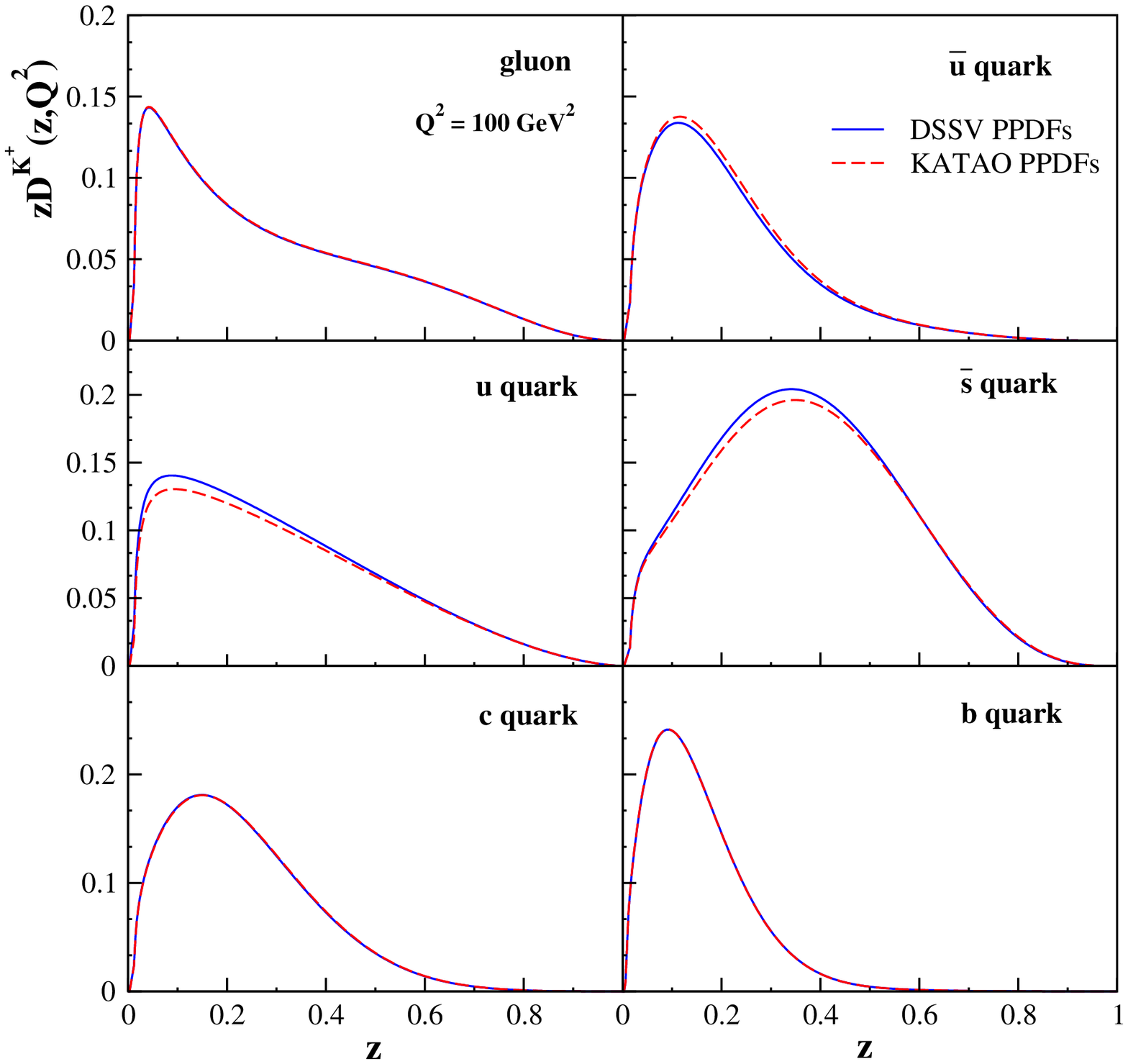}}
\vspace{-0.4cm}
\caption{Comparison of extracted kaon fragmentation functions at $Q^{2} = 100$ GeV$^2$ using KATAO \cite{Khorramian:2010qa}
and DSSV \cite{deFlorian:2009vb} PPDFs at NLO.}
\vspace{-0.5cm}
\label{KDSSV-KATAO}
\end{figure}
%
\begin{figure}[t!]
\vspace{-0.4cm}
\centerline{\includegraphics[width=0.42\textwidth,angle=-90]{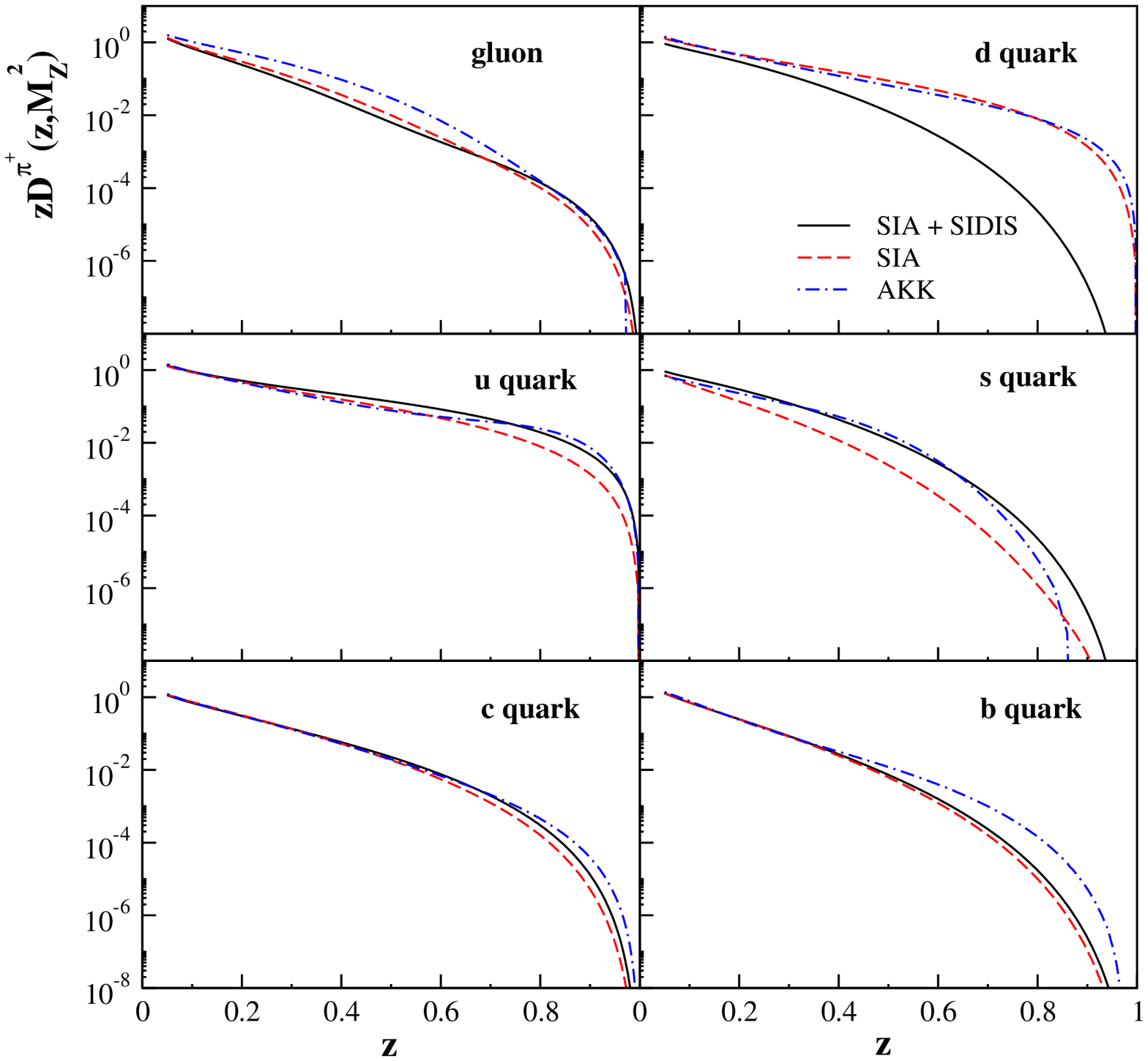}}
\vspace{-0.3cm}
\caption{Comparison of calculated $\pi ^+$ FFs at
$Q^2= M_Z^{2}$ from fitting on SIA data~\cite{aleph91,delphi91,delphi91-2,opal91,Lees:2013rqd,sld91,tasso34_44,tpc29,Leitgab:2013qh,topaz58} with (solid lines)
and without (dashed lines) SIDIS data~\cite{Hermes05,Alekseev:2010ub,Alekseev:2009ac} at NLO.
The results are compered with AKK~\cite{Albino:2008fy} (dot-dashed lines) too.}
\label{pcomparison}
\centerline{\includegraphics[width=0.42\textwidth,angle=-90]{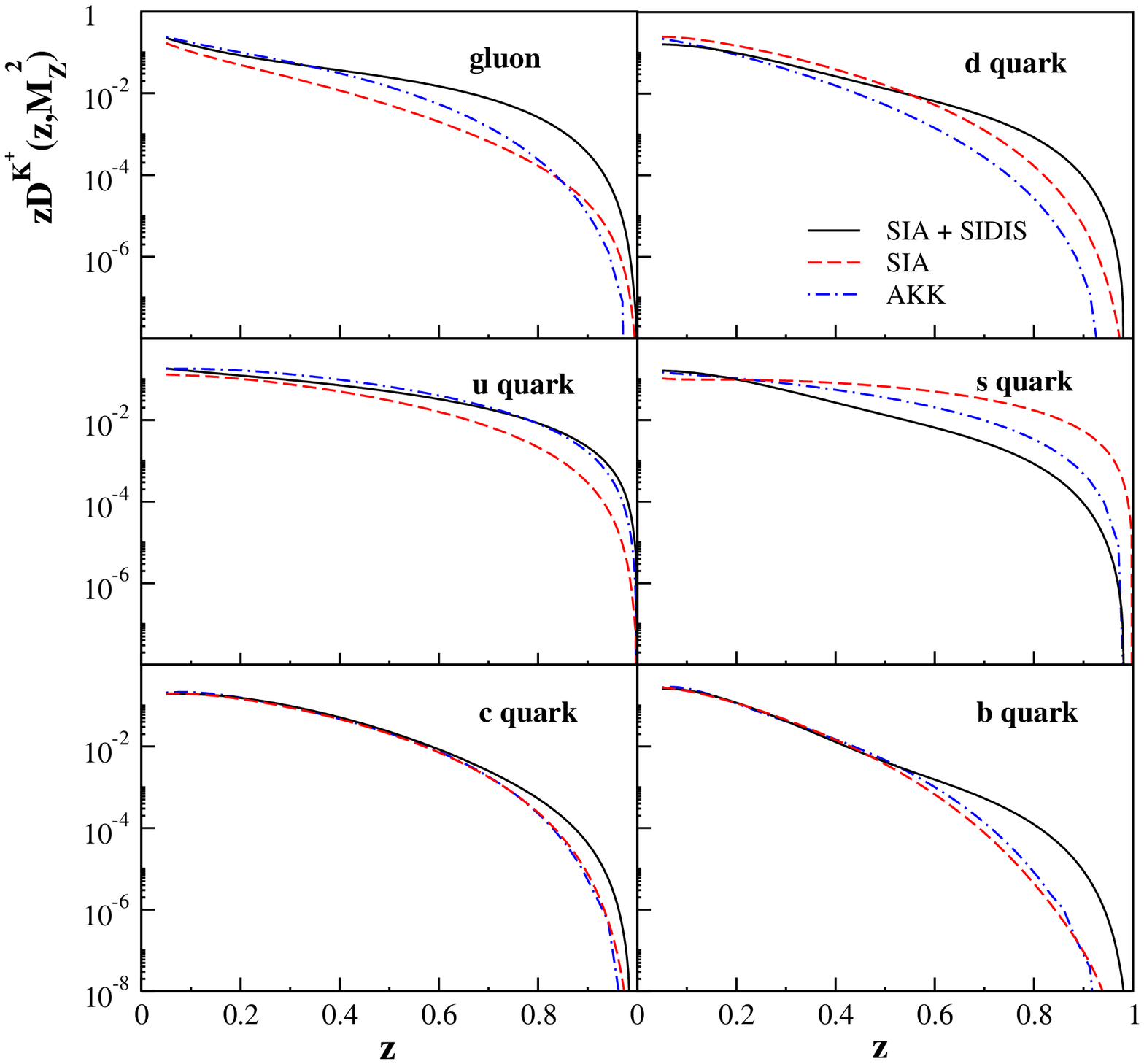}}
\vspace{-0.4cm}
\caption{ Comparison of calculated $K^+$ FFs at
$Q^2= M_Z^{2}$ from fitting on SIA data~\cite{aleph91,delphi91,delphi91-2,opal91,Lees:2013rqd,sld91,tasso34_44,tpc29,Leitgab:2013qh,topaz58} with (solid lines)
and without (dashed lines) SIDIS data~\cite{Hermes05,Alekseev:2010ub,Alekseev:2009ac} at NLO.
The results are compered with AKK~\cite{Albino:2008fy} (dot-dashed lines) too.}
\vspace{-0.5cm}
\label{kcomparison}
\end{figure}
%
\begin{figure*}
\vspace{+0.4cm}
\centerline{\includegraphics[width=0.5\textwidth,angle=-90]{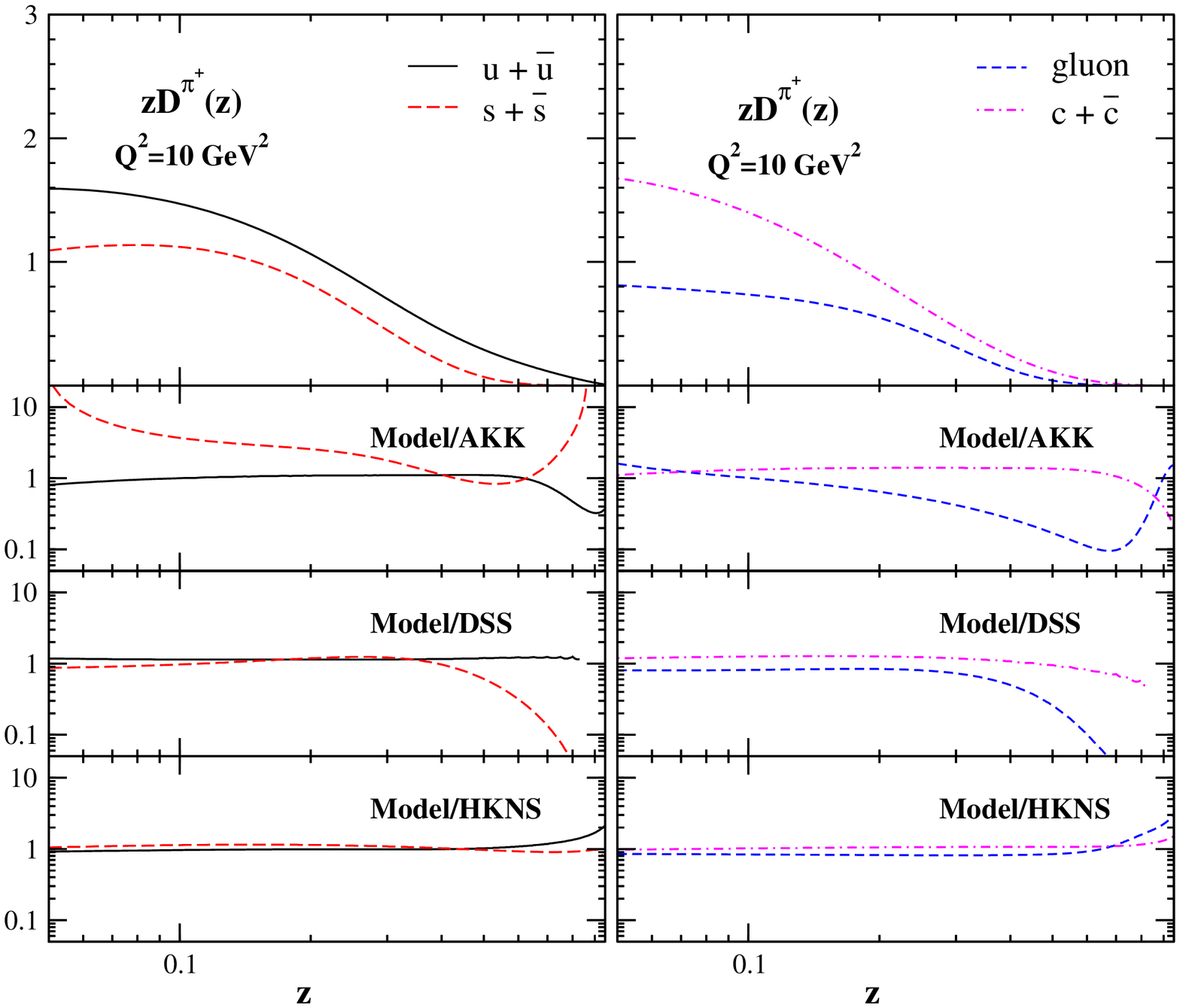}}
\vspace{-0.7cm}
\caption{Upper panels: fragmentation densities for $\pi ^+$
at $Q^2=10$~GeV$^2$ at NLO. Rest panels: ratios of our fragmentation densities to the ones of HKNS, DSS and AKK~\cite{Albino:2008fy,deFlorian:2007aj,Hirai:2007cx}.}
\label{ratiopion10}
\centerline{\includegraphics[width=0.5\textwidth,angle=-90]{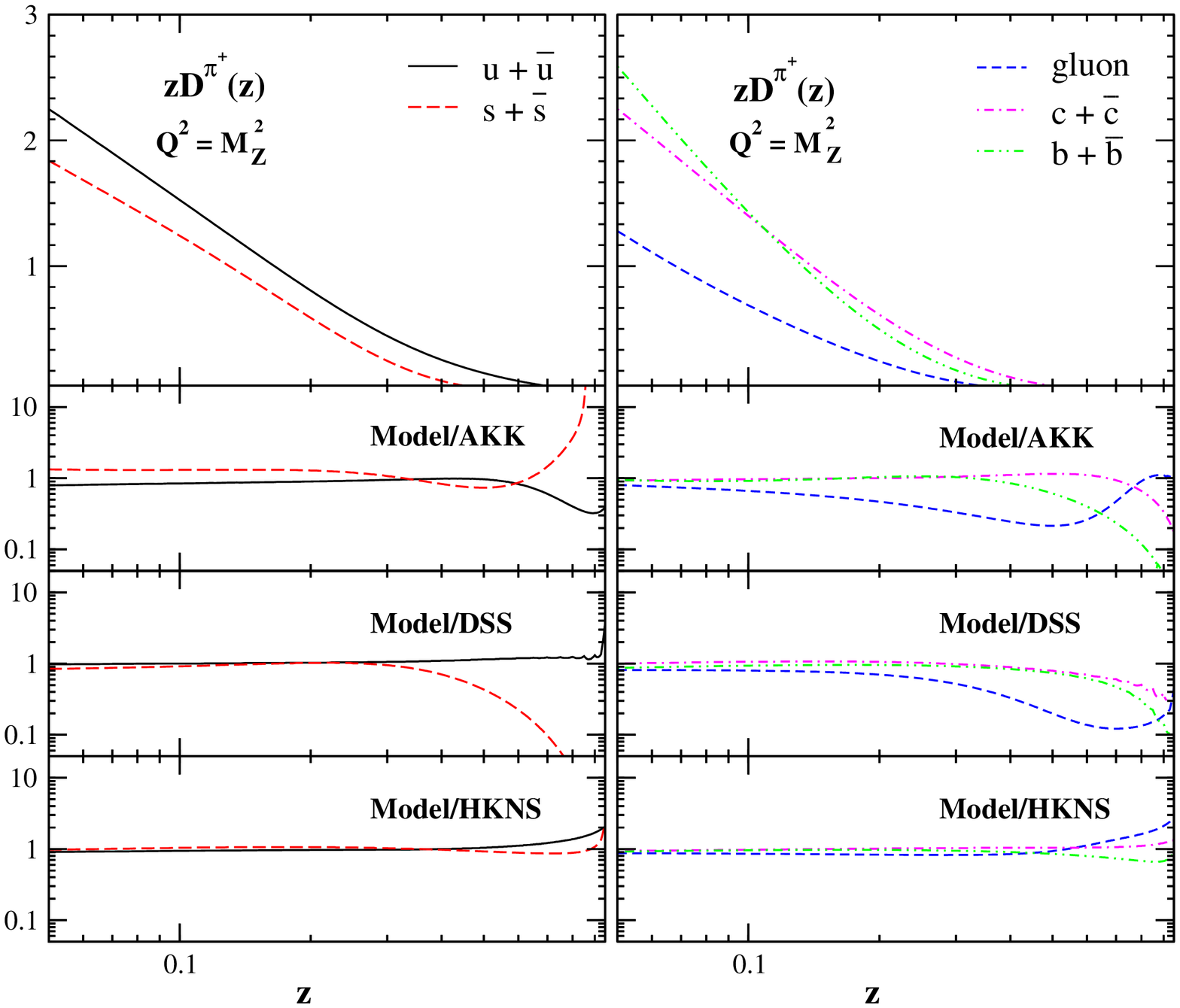}}
\vspace{-0.8cm}
\caption{ Upper panels: fragmentation densities for $\pi^+$
at $Q^2=M_Z^{2}$ at NLO. Rest panels: ratios of our fragmentation densities to the ones of HKNS, DSS and AKK~\cite{Albino:2008fy,deFlorian:2007aj,Hirai:2007cx}.}
\label{ratiopionMz}
\vspace{-0.4cm}
\end{figure*}
%
\begin{figure*}
\vspace{+0.4cm}
\centerline{\includegraphics[width=0.5\textwidth,angle=-90]{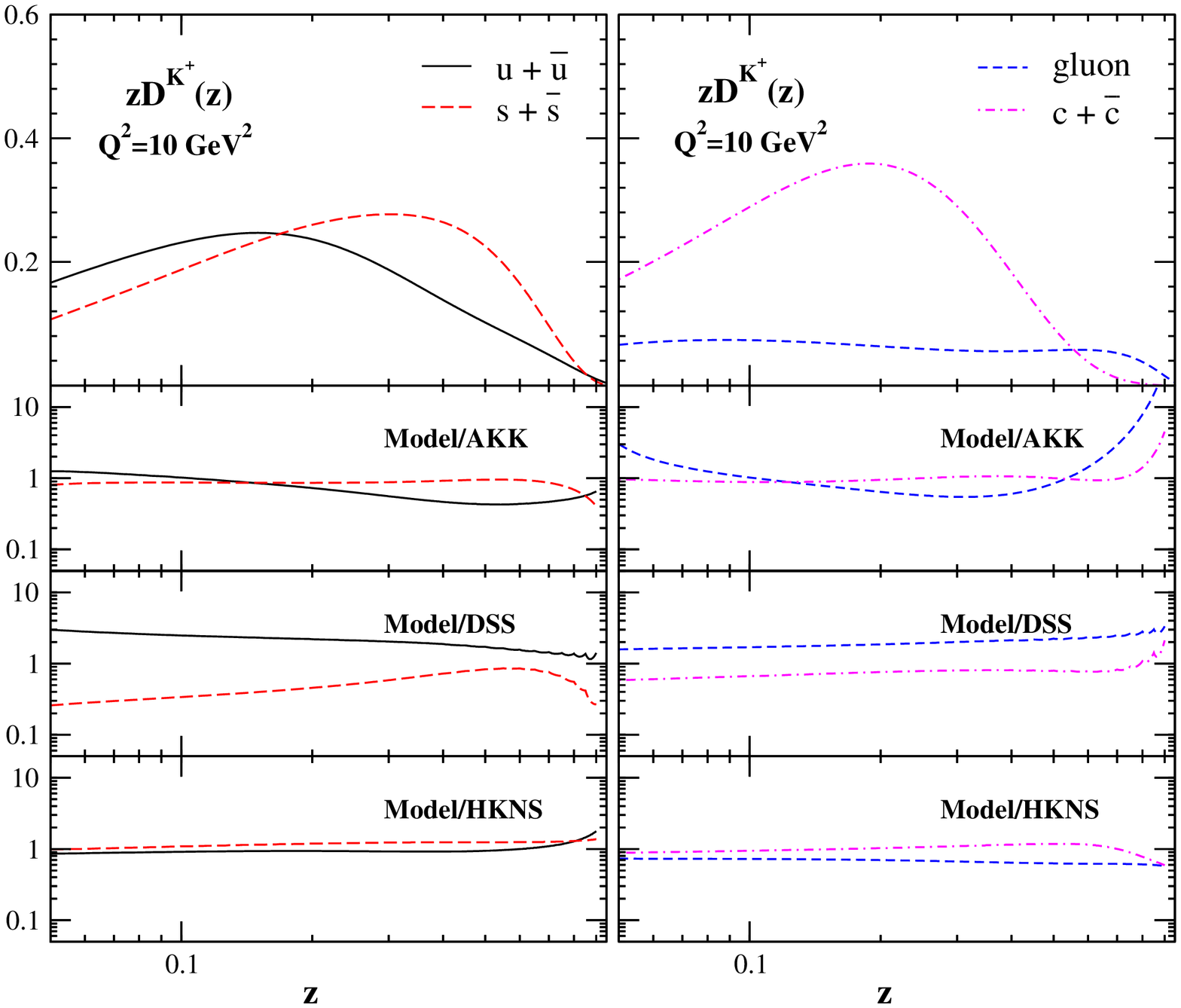}}
\vspace{-0.8cm}
\caption{Upper panels: fragmentation densities for $K^+$ at
$Q^2=10$~GeV$^2$ at NLO. Rest panels: ratios of our fragmentation densities to the ones of HKNS, DSS and AKK~\cite{Albino:2008fy,deFlorian:2007aj,Hirai:2007cx}.}
\label{ratiokaon10}
\centerline{\includegraphics[width=0.5\textwidth,angle=-90]{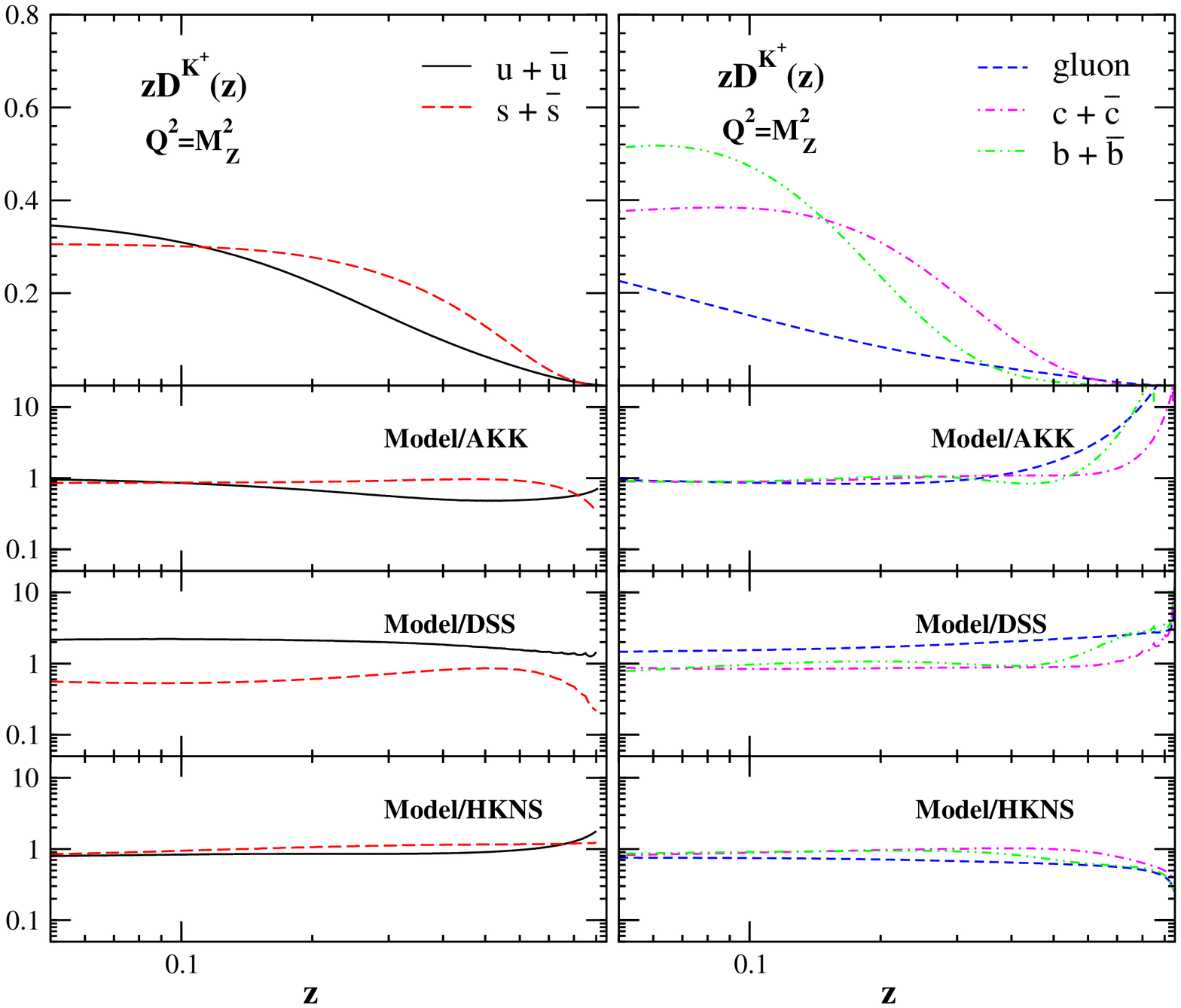}}
\vspace{-0.8cm}
\caption{ Upper panels: fragmentation densities for $K^+$ at
$Q^2=M_Z^{2}$ at NLO. Rest panels: ratios of our fragmentation densities to the ones of HKNS, DSS and AKK~\cite{Albino:2008fy,deFlorian:2007aj,Hirai:2007cx}.}
\label{ratiokaonMz}
\vspace{-0.4cm}
\end{figure*}
\hspace{-0.65cm}To show how different choices of PPDFs from different analysis affect the results, we applied
one of the most accurate PPDFs, i.e. KATAO PPDFs \cite{Khorramian:2010qa} which are obtained from a global analysis of DIS data. The comparison of extracted pion and kaon fragmentation functions by including KATAO \cite{Khorramian:2010qa} and DSSV \cite{deFlorian:2009vb} PPDFs are shown in Figs.~\ref{pDSSV-KATAO} and \ref{KDSSV-KATAO} at NLO.
As it is seen the differences are small and negligible and these differences become smaller by increasing the energy scaling.
Therefore, the different choices of PPDFs do not change our result considerably.\\
Since we would like to present how much the asymmetry SIDIS
data are effective for determination of FFs, in Figs~\ref{pcomparison} and \ref{kcomparison} the FFs for different flavors are presented in two
cases at $\mu^2=M_Z^{2}$. In the first case we determine FFs by fitting on the single-inclusive electron-positron annihilation (SIA) and also SIDIS asymmetry data. According to the last section, we assume asymmetry between valence or favored fragmentation functions
and unfavored fragmentation functions for both pion and kaon because the possibility of
$\pi^+/K^+$-production from valence or favored quarks is more than sea
or unfavored quarks. Moreover, SIDIS data help us
to specify the difference between the quark and anti-quark
distributions in the nucleon considering outgoing produced hadrons
 which is not possible in fully inclusive experiments.\\
In the second case we calculate FFs by fitting just on the single-inclusive electron-positron annihilation (SIA) data.
Since we omit asymmetry SIDIS data from our fit, the symmetry between the quark and anti-quark is assume in this case
\begin{eqnarray}
\label{ff-kaon2}
&D_u^H(z,\mu_{0}^{2})& = D_{\bar{u}}^H(z,\mu_{0}^{2}),\nonumber\\
&D_d^H(z,\mu_{0}^{2})& = D_{\bar{d}}^H(z,\mu_{0}^{2}),\nonumber\\
&D_s^H(z,\mu_{0}^{2})& = D_{\bar{s}}^H(z,\mu_{0}^{2}).
\end{eqnarray}
According to Figs~\ref{pcomparison} and \ref{kcomparison}, the SIDIS data are effective on different partons of FFs.
Also our results are compered with AKK model in these Figs.\\
We also present the $\pi^+$ and $K^+$ fragmentation densities at the scales $\mu^2=10$~GeV$^2$
and $\mu^2=M_Z^{2}$ and  the ratios of our fragmentation densities
to the ones presented by HKNS, DSS and AKK~\cite{Albino:2008fy,Hirai:2007cx,deFlorian:2007aj} are shown in
 Figs.~\ref{ratiopion10}, \ref{ratiopionMz}, \ref{ratiokaon10} and \ref{ratiokaonMz}.
 According to these figures, our FF models  densities are different in comparison with other models at large $z$
and it is not unexpected because according to Figs.~\ref{pcomparison} and \ref{kcomparison}, the SIDIS data impression
on the FFs at large $z$ is more than small $z$ then the ratios in Figs.~\ref{ratiopion10}, \ref{ratiopionMz}, \ref{ratiokaon10} and \ref{ratiokaonMz} are mush better agreement at small $z$. Although when $Q^2$  increases the difference between models decreases.\\
Since, unlike our assumption, the light-quarks functional forms are separated in the AKK analysis due to fully flavor separated OPAL data,
then the difference between our FFs results and AKK results is more than DSS and HKNS  in Figs.~\ref{ratiopion10}, \ref{ratiopionMz}, \ref{ratiokaon10} and \ref{ratiokaonMz}.
These data which are not used in the all analysis, such as HKNS, are more difficult to appreciate within purturbative  QCD beyond the LO.
Then we just use of untagged OPAL data and also the light-quarks functions are not separated in our analysis such as DSS and HKNS analysis.
\subsection{Quadratic behavior of $\Delta\chi^{2}$}
\begin{figure}[h!]
\hspace{-1.cm}
\centerline{\includegraphics[width=0.42\textwidth,angle=-90]{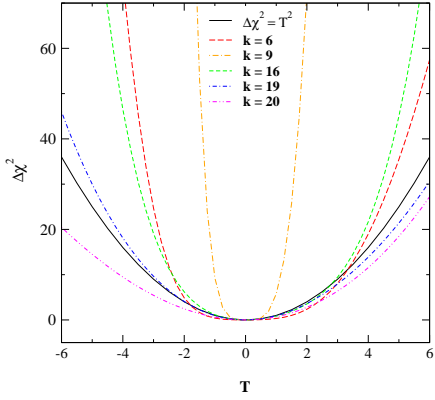}}
\vspace{-0.7cm}
\caption{Examples of pion $\Delta\chi^2$ deviations from
the expected quadratic behavior $\Delta\chi^2=T^2$ for
random sample eigenvector directions.}
\label{Deltachi2p}
\end{figure}

\begin{figure}[h!]
\vspace{-0.4cm}
\hspace{-1.cm}
\centerline{\includegraphics[width=0.42\textwidth,angle=-90]{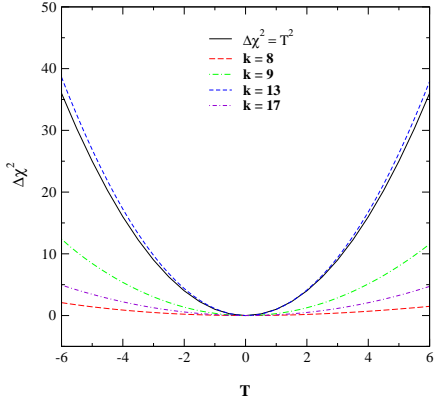}}
\vspace{-0.7cm}
\caption{Examples of kaon $\Delta\chi^2$ deviations from the
expected quadratic behavior $\Delta\chi^2=T^2$ for
random sample eigenvector directions.}
\label{Deltachi2k}
\vspace{0.cm}
\end{figure}

According to the Hessian method which is discussed in \ref{subsec:hessian method} we want to indicate if $\Delta\chi^{2}$ shows the assumed quadratic behavior on the parameters from the best fit.
To explore this further, we present the dependance
of $\Delta\chi^{2}$ global along some random samples of eigenvector
directions to illustrate the deviations of the $\Delta\chi^{2}$ function from the expected quadratic dependence.
As can be seen, to exhibit the quadratic approximation in Eq.~\ref{eq:hessian}, Fig.~\ref{Deltachi2p} and Fig.~\ref{Deltachi2k}
are presented to show the pion and kaon $\Delta\chi^{2}$ global along some random samples of eigenvector directions.
%
Since the variation range of fitted parameters is correlated, here only one of the parameters
is varied. Fig.~\ref{Deltachi2p} presents pion $\Delta\chi^{2}$ along some random samples of eigenvector
directions and eigenvalues, $k=6,9,16,19$ and $20$.
In this figure the curve with the eigenvectors direction $k=19$ for pion
shows the most idealistic quadratic behavior and some other curves with $k=6,9$ a deviation from the ideal parabolic behavior curve $\Delta\chi^{2}=T^{2}$. Also Fig.~\ref{Deltachi2k} presents kaon $\Delta\chi^{2}$ along some random samples of eigenvector
directions and eigenvalues, $k=8,9,13$ and $17$.
To have a best fit we omit the term $[1-e^{-\gamma_iz}]$
for heavy partons $c/\bar{c}$ and $b/\bar{b}$ in global analysis of kaon (see Tables~\ref{tab:lokaonpara} and \ref{tab:nlokaonpara})
and it improves the quadratic behavior of $\Delta\chi^2$.
More details about the kaon parameters are explained in Sec.~\ref{subsec:threeB}.
\begin{figure}[h!]
\hspace{0.5cm}
\centerline{\includegraphics[width=0.42\textwidth,angle=-90]{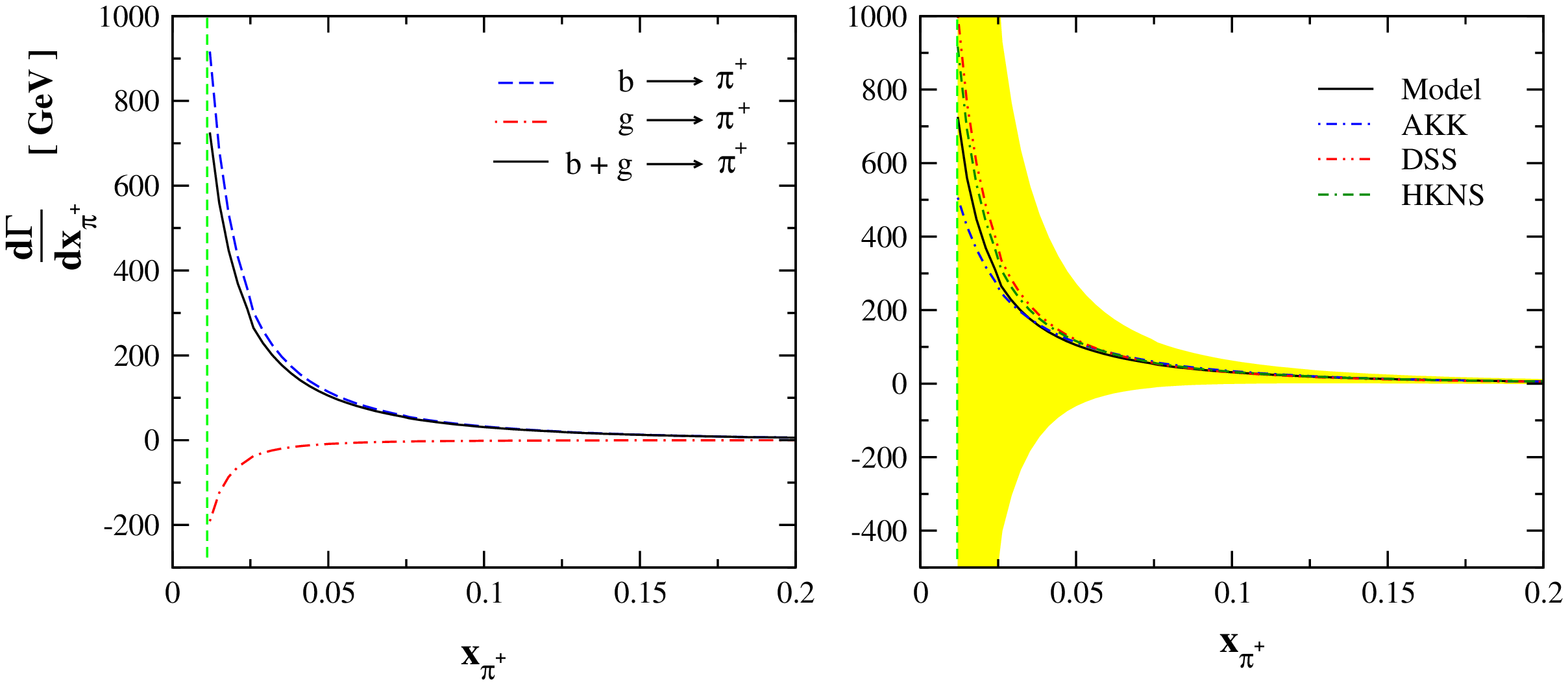}}
\vspace*{-3.cm}
\caption{Left panel: Energy distribution of $\pi^+$-meson in top decay considering the fragmentation contribution of b-quark (dashed line) and gluon (dot-dashed line) into the $\pi^+$ and the total contribution (solid line) at $\mu_F=m_t$. Right panel: Energy distribution and its uncertainty of $\pi^+$ considering the FFs obtained by our model, AKK, DSS and HKNS~\cite{Albino:2008fy,deFlorian:2007aj,Hirai:2007cx}.}
\label{toptopion}
\vspace{-0.7cm}
\hspace{0.5cm}
\centerline{\includegraphics[width=0.42\textwidth,angle=-90]{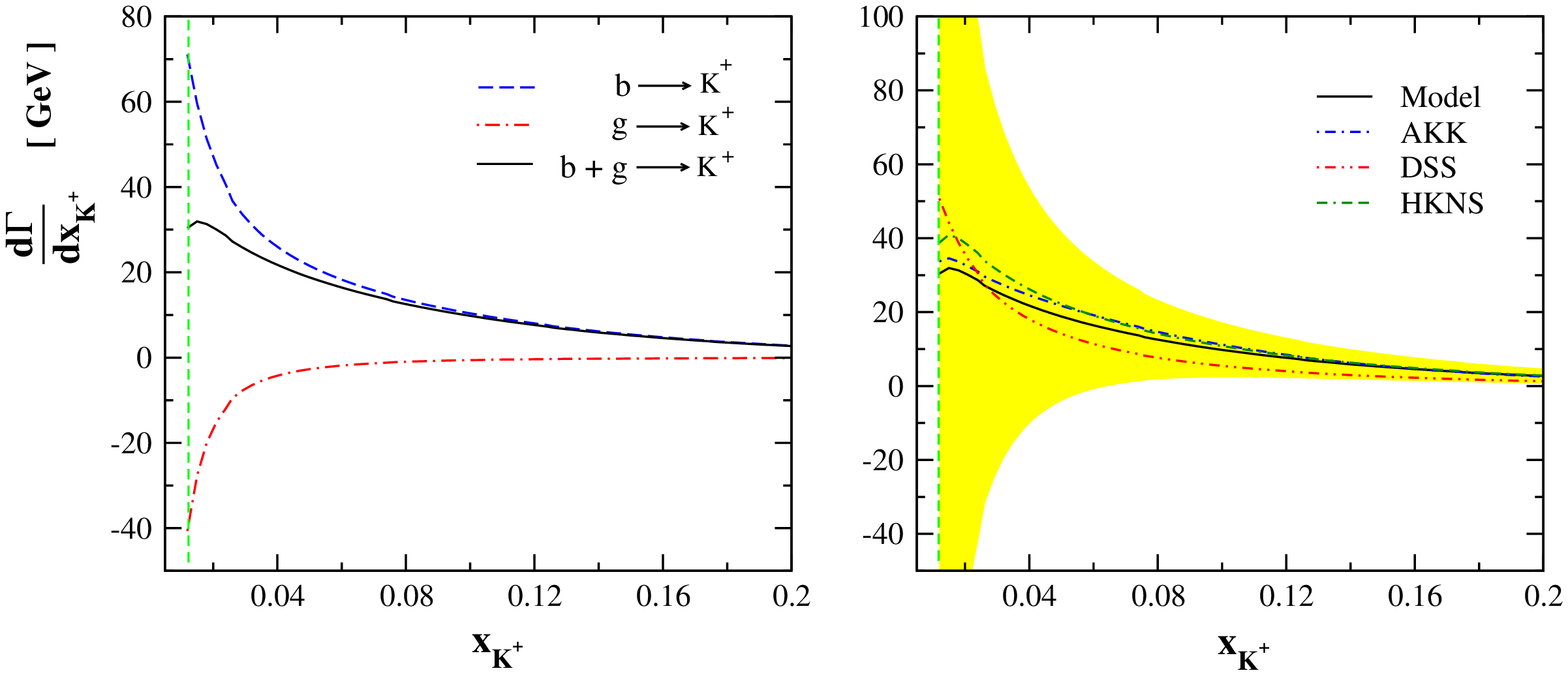}}
\vspace*{-0.7cm}
\caption{ Left panel: Energy distribution of $K^+$-meson in top decay considering the fragmentation contribution of b-quark (dashed line) and gluon (dot-dashed line) into the $K^+$ and the total contribution (solid line) at $\mu_F=m_t$. Right panel: Energy distribution and its uncertainty of $K^+$ considering the FFs obtained by our model, AKK, DSS and HKNS~\cite{Albino:2008fy,deFlorian:2007aj,Hirai:2007cx}.}
\label{toptokaon}
\end{figure}
%
\subsection{Energy spectrum of light mesons in top-quark decay}
Now, by having the pion and kaon fragmentation functions in every scale, we make our phenomenological prediction for  energy spectrum of light mesons in top decay.
Therefor we adopt from Ref.~\cite{Nakamura:2010zzi} the input parameter values
$G_F = 1.16637\times10^{-5}$~GeV$^{-2}$,
$m_t = 172.0$~GeV, and
$m_{W^+}=80.399$~GeV.
In Figs.~\ref{toptopion} and \ref{toptokaon}, we show our predictions for the size of the
NLO corrections and their uncertainties, by comparing the relative importance of the $b\to \pi^{+}/K^{+}$ (dashed line)
 and $g\to \pi^{+}/{K^+}$ (dot-dashed line) fragmentation,  on a  logarithmic scale. As it is seen
the gluon fragmentation leads to an  appreciable reduction in decay rate at low-$x_H$ region
and for higher values of $x_H$,  the $b\to H$ contribution is dominant.
The mass of  light meson is responsible for the appearance of the threshold at $x_H=2m_H/(m_t^2-m_W^2)$.
For comparison, we also show the energy spectrum of light mesons in top decay
using the FFs obtained by AKK, DSS and HKNS collaborations in Figs.~\ref{toptopion} and \ref{toptokaon}.
\section{Conclusions}
\label{sec:six}
We have determined the
non-perturbative parton fragmentation functions for pion and kaon at
LO and NLO approximation from global analysis of single-inclusive
electron-positron annihilation $e^{+}e^{-}\rightarrow (\gamma ,Z)\rightarrow H+X$ and double spin asymmetry
from semi-inclusive deep inelastic scattering data $A_{1}^{N,H}$, $\vec{l}(l) + \vec{N}
\rightarrow l^{'}(l^{'}) + H +X$. Our
analysis was based on zero mass variable flavor number scheme (ZM-VFNS) where all quarks are treated as massless particles.
Our new  parameterization form covers a wide kinematic range $z$ because of the extra term $[1-e^{-\gamma_i z}]$ which controls
medium $z$ region and improves the accuracy of the global fit. Figs.~\ref{pion+data},~\ref{kaon+data},~\ref{BBdata},~\ref{pi} and \ref{ki} show the comparison of our model with SIA and double spin asymmetry SIDIS experimental data and indicate that our model is successful.
We determined the FFs of gluon and light-quarks at the initial scale $\mu_{0}^{2}=1$~GeV$^2$ and the FFs of heavy quarks at $\mu_{0}^2=m_c^{2}$ and $\mu_{0}^{2}=m_b^{2}$.
Evaluation determined by the DGLAP equations.
The theoretical results of $b$ heavy quark for pion in our model and other models such as HKNS and DSS \cite{Hirai:2007cx,deFlorian:2007aj} deviate from the SLD and DELPHI data at large $z$ and any deviance between theory
and experimental data occurs a large $\chi^{2}$. In comparison with other group we applied, for the first time, spin asymmetry data ($A_{1}^{N,H}$) in the global
analysis of the fragmentation functions and
the energy scales which are reported for
the SIDIS experimental data are low energy scales which are
usually smaller than the $e^{+}e^{-}$ annihilation scales (see Tables \ref{tab:exppionLO}, \ref{tab:exppionNLO}, \ref{tab:expkaonLO} and \ref{tab:expkaonNLO}).
On the other hand adding the SIDIS data in a global
fit leads us to test the universality of parton fragmentation functions so that
the results are in a good agreement with the FFs of other models.
We also used one of the most accurate polarized and unpolarized parton distribution functions, i.e. NLO DSSV for polarized PDFs and NLO KKT12 for unpolarized
PDFs. Using PDFs to determine FFs both indicates the universality of PDFs and is a good test for perturbative QCD analysis.
We can also apply fragmentation functions to determine parton distribution  into the proton, deuteron and neutron and
show that parton densities do not depend on the corresponding cross sections and are universal.
At the end we used the
fragmentation functions to predict the energy spectrum of $\pi^+$
and $K^+$ mesons produced in top-quark decay. Comparison of our results for  pion and kaon energy spectrum with other models shows  that the fragmentation functions are universal, see Figs.~\ref{toptopion} and \ref{toptokaon}.

\acknowledgments
We warmly acknowledge F. I. Olness, S. J. Brodsky
and G. Kramer for valuable discussions, critical remarks and reading the manuscript. We appreciate E. Christova and M. Leitgab for useful suggestions and comments.
A. N. K. thanks the SITP (Stanford Institute for Theoretical Physics) for partial support and the Physics Department of SMU (Southern Methodist University) for their hospitality during the completion of this work.
A. N. K. and M. M. thank the CERN TH-PH division for its hospitality where a portion of this work was performed. We thank the School of Particles and Accelerators, Institute for Research in Fundamental Sciences (IPM) for financial support.

\section*{Appendix: FORTRAN-code }
\label{App}
A \texttt{FORTRAN} package containing our unpolarized fragmentation functions for pion and
kaon at LO and NLO can be found in \texttt{http://particles.ipm.ir/links/QCD.htm}~\cite{Program-summary} or
obtained via e-mail from the authors. These functions are interpolated
using cubic splines in $Q^{2}$ and a linear interpolation in $\log\,(Q^{2})$.
The
package includes an example program to illustrate the use of the routines.

\end{document}